\documentclass[12pt, a4paper]{article}

\pdfoutput=1

\usepackage{amsmath}
\usepackage{amsfonts}
\usepackage{amssymb}
\usepackage{cite}
\usepackage{multicol,multirow}
\usepackage{graphicx, physics, subcaption}
\graphicspath{{figs/}}
\usepackage{graphics, epsfig}
\usepackage{epsf}
\usepackage{epstopdf}

\usepackage[usenames,dvipsnames]{color}

\usepackage{colortbl}
\definecolor{summersky}{cmyk}{0.71,0.33,0,0.14}
\definecolor{flamingo}{cmyk}{0,0.51,0.71,0.14}
\definecolor{rp}{cmyk}{0.2, 1, 0.6, 0}
\definecolor{pacificblue}{cmyk}{0.95,0.3,0, 0.19}
\definecolor{gray60}{cmyk}{0.4,0.4,0,0.8}

\numberwithin{equation}{section}

\usepackage[colorlinks,bookmarks=false, hypertexnames=true]{hyperref}
\hypersetup{
	breaklinks=true,
	pdfstartview={FitH},    
	colorlinks=true,       
	linkcolor=flamingo,          
	citecolor=summersky,        
	filecolor=magenta,      
	urlcolor=pacificblue,           
	anchorcolor=green,      
	linktocpage=true
}

\newcommand{\nc}{\newcommand}

\newcommand{\sigmaE}{\sigma^{_E}}
\newcommand{\sigmaB}{\sigma^{_B}}
\newcommand{\bdsigmaE}{{\boldsymbol \sigma}^{_E}}
\newcommand{\bdsigmaB}{{\boldsymbol \sigma}^{_B}}

\newcommand{\DE}{D_{_E}}
\newcommand{\DB}{D_{_B}}

\nc{\ba}{\begin{eqnarray}}
\nc{\ea}{\end{eqnarray}}

\nc{\calR}{{\cal{R}}}
\nc{\calP}{{\cal{P}}}
\nc{\cN}{ {\cal{N}} }

\def\bfx{{\bf x}}

\newcommand{\Mp}{M_{_{\rm Pl}}}
\nc{\Mpt}{{\rm {M^2_{_P}}}}

\usepackage[normalem]{ulem}
\newcommand\redsout{\bgroup\markoverwith{\textcolor{red}{\rule[0.4ex]{2pt}{2pt}}}\ULon}

\begin{document}


\begin{center}

{\bf  Stochastic  Effects in Axion Inflation and Primordial Black Holes Formation}
\\[0.5cm]

{Alireza Talebian \footnote{talebian@ipm.ir}, 
Amin Nassiri-Rad \footnote{amin.nassiriraad@ipm.ir}, 
Hassan Firouzjahi \footnote{firouz@ipm.ir},
}
 
 {\small \textit{School of Astronomy, Institute for Research in Fundamental Sciences (IPM) \\ P.~O.~Box 19395-5531, Tehran, Iran
}}\\

\today
\end{center}

\hrule
\begin{abstract}
\label{abstract}
We revisit the model of axion inflation in the context of  stochastic inflation and investigate the effects of  the stochastic noises associated to the electromagnetic fields. Because of the parity violating interaction, one polarization of the gauge field is amplified inducing large curvature perturbation power spectrum.  Taking into account the stochastic kicks arising from the short modes at the time of horizon crossing  we obtain the corresponding Langevin equations for the long modes of the  electromagnetic and axion fields. It is shown that a mean-reverting process governs the dynamics of the electromagnetic fields such that the tachyonic growth of the gauge fields is balanced by the diffusion forces. As the instability parameter grows towards the end of inflation, the large curvature perturbations induced from gauge field perturbations  lead to copious production of small mass primordial black holes (PBHs). It is shown that the produced PBHs follow a Gaussian statistics.  Imposing the observational constraints on PBHs formation relaxes the previous bounds on the instability parameter by about fifty percents.
 \\
\end{abstract}
\vspace{0.3cm}
\hrule
\tableofcontents
\section{Introduction}
\label{sec:intro}

Inflation is a cornerstone of early universe cosmology which is well supported by cosmological observations \cite{Planck:2018jri}. Simplest models of inflation 
are based on a single scalar field, the inflaton field, which rolls on top of a nearly flat potential. These simple scenarios predict that the curvature perturbations on superhorizon scales to be nearly scale invariant, nearly adiabatic and nearly Gaussian which are consistent with observations. However, despite its successes, inflation is still a phenomenological paradigm looking for a deeper theoretical understanding. Among key questions are what is the nature of the inflaton field or what mechanism keeps the potential flat enough to sustain a long enough period of inflation to solve the flatness and the horizon problems.

One of the well-motivated proposal to protect the potential against the ultra-violate (UV) corrections and to keep it nearly flat is  to assume that inflaton is a Pseudo-Nambu-Goldstone Boson (PNGB) field
\cite{Freese:1990rb,Adams:1992bn,ArkaniHamed:2003wu,Kim:2004rp,Dimopoulos:2005ac,Easther:2005zr,McAllister:2008hb,Kaloper:2008fb,Flauger:2009ab,Anber:2009ua}. PNGBs, like the axion, are pseudo-scalar fields which arise whenever a global symmetry is spontaneously broken. Pseudo-scalar fields with axial symmetry are very common in particle physics, and enjoy a shift symmetry $\phi\rightarrow \phi + const.$ which is broken either explicitly or by quantum effects.  In the limit of approximate symmetry, the corrections to the slow-roll parameters are controlled by the smallness of the symmetry breaking. Inflationary scenarios where a  pseudo-scalar is identified as the inflaton or a spectator field affecting the inflationary dynamics have been widely discussed in the past \textit{e.g.} see Refs.~\cite{Freese:1990rb,Adams:1992bn,Lue:1998mq,Kim:2004rp,Dimopoulos:2005ac,McAllister:2008hb,Kaloper:2008fb,Kaloper:2011jz, Pajer:2013fsa}. 
Natural inflation \cite{Freese:1990rb} is among the first models of axion inflation 
in which the shift symmetry is broken down to a discrete subgroup $\phi \rightarrow \phi + 2\pi f$, resulting in a periodic potential
\begin{equation}
\label{Vnp}
V_{\mathrm{np}}(\phi) = \Lambda^4 \Big[1-\cos\big(\dfrac{\phi}{f}\big)\Big] \,,
\end{equation}
in which $f$, known as  the axion decay constant, has the dimensions of mass.  In these models, the inflaton is coupled to a $U(1)$ gauge field $A_\mu$ via the interaction of the form
\begin{equation}
\label{int}
\mathcal{L}_{\mathrm{int}} = - \frac{\alpha}{4 f}\, \phi \, F^{\mu\nu}\tilde{F}_{\mu\nu} \,,
\end{equation}
where $F_{\mu\nu}=\partial_\mu A_\nu-\partial_\nu A_\mu$ is the field strength, and $\tilde{F}^{\mu \nu} \equiv \frac{\epsilon^{\mu \nu \alpha \beta}}{2\sqrt{-g}}F_{\alpha \beta}$ is its dual with $\epsilon^{0123}=1$. The strength of the interaction is controlled by $f$ and the dimensionless parameter $\alpha$. 

This type of parity violating interaction causes  interesting cosmological effects including enhancing the scalar power spectra and  non-Gaussianity \cite{Barnaby:2010vf, Meerburg:2012id}, chiral gravitational waves \cite{Adshead:2012kp,Maleknejad:2011jw,Dimastrogiovanni:2012ew,Adshead:2013nka,Namba:2013kia,Maleknejad:2016qjz, Adshead:2016omu, Obata:2016xcr,Dimastrogiovanni:2016fuu, Dimastrogiovanni:2018xnn, Gorji:2020vnh, Salehian:2020dsf} at CMB \cite{Sorbo:2011rz,Barnaby:2012xt,Namba:2015gja} and interferometer \cite{Cook:2011hg,Barnaby:2011qe,Domcke:2016bkh} scales. Moreover, the interaction  \eqref{int} has important implications for the primordial black holes (PBHs) 
formation \cite{Linde:2012bt,Cheng:2016qzb}. 
As pointed out in Ref.~\cite{Anber:2006xt}, since the interaction  \eqref{int} violates parity, the rolling of inflaton causes one polarization of the gauge field to become tachyonic.  The tachyonic growth of the gauge field quanta 
then backreacts on the inflaton field itself via {inverse decay}: $\delta A + \delta A \rightarrow \delta\varphi$, causing the enhancement of the scalar power spectra and other interesting effects as mentioned above.

Stochastic inflation is an IR effective field theory  to study the dynamics of the superhorizon perturbations while the small scales are continuously stretched to superhorizon scales acting as the  source of the classical noises  
\cite{Starobinsky:1986fx,Sasaki:1987gy, Nambu:1988je, Nakao:1988yi, Rey:1986zk, Starobinsky:1994bd}. The stochastic effects of the gauge fields perturbations during inflation have been studied in \cite{Talebian:2019opf, Talebian:2020drj,Talebian:2021dfq}, see also 
\cite{Fujita:2017lfu}.  It was shown in these works that the stochastic dynamics of the electromagnetic perturbations can have non-trivial effects on the physical predictions. Specifically, in models of inflation involving the gauge fields one typically imposes the conditions that the electromagnetic backreactions to be under control, for example the energy density associated to electromagnetic fields to be always smaller than the inflaton energy density. It was shown in \cite{Talebian:2019opf, Talebian:2020drj,Talebian:2021dfq} 
that the stochastic nosies associated to electromagnetic perturbations  modify the contributions of the electromagnetic fields in power spectra or the amplitudes of the primordial magnetic fields. It is concluded, among other things,  that the stochastic effects can relax the backreaction constraints yielding to a modification of the model parameters such as the gauge kinetic coupling.  Motivated by these results, we revisit the scenario of axion inflation using the formalism of stochastic inflation. 
We show that the  results for the curvature perturbation power spectrum induced
from the gauge fields perturbations obtained from stochastic formalism are overall consistent with the previous results in the literature. However,  the stochastic  effects modify the allowed parameter space of the model. In addition, we provide new insights for the backreaction effects  in the context of stochastic formalism.

The paper is organized as follows. In section \ref{Ratra-Model}  we review  the setup of axion inflation in conventional approach in 
the absence of stochastic noises.  In section \ref{Eom-long} we employ stochastic formalism to the model of axion inflation and solve the Langevin equations associated to the electric and magnetic fields.  In section \ref{backreaction-sec} we look at the backreaction effects in the presence of stochastic noises while
in section \ref{Subsec.Phi} we study the scalar power spectrum using the stochastic $\delta N$ formalism. In section \ref{sec:PBH} we study the  PBHs formation in this setup and their cosmological constraints 
followed by Summary and Discussions in section \ref{sec:Discussion}. Various technicalities regarding the noises and their correlations and technical applications of stochastic $\delta N$ formalism are relegated to appendices 
\ref{scalar-mode}-\ref{appendix:PDF}.

\section{Overview of the Model}
\label{Ratra-Model}

In this section we review the setup of axion inflation and present the  results in literature obtained in the absence of stochastic effects. 

The model  consists of a pseudo-scalar inflaton field $\phi$ interacting with a $U(1)$ gauge field $A_\mu$ given by the following action 
\begin{eqnarray}
		\mathcal{S} = \int \mathrm{d}^4 x \, \sqrt{-g} \, \bigg[ \dfrac{\Mp^2}{2} R 
		- \dfrac{1}{2} g^{\mu \nu} \partial_{\mu} \phi  \partial_{\nu} \phi - V(\phi) 
		-\frac{1}{4}F_{\mu\nu}F^{\mu\nu}+ \mathcal{L}_{\mathrm{int}}\bigg] \,,
	\label{action}
\end{eqnarray}
in which $ \mathcal{L}_{\mathrm{int}}$ is the  parity violating interaction \eqref{int},  
$\Mp$ is the reduced Planck mass and $R$ is the Ricci scalar. The metric $g_{\mu\nu}$represents a spatially flat FLRW space-time, 
\begin{align}
	\label{metric}
\dd s^2 &= -\dd t^2 + a^2(t) \ \dd \bfx \cdot \dd \bfx 
\end{align}
in which $t$ is the cosmic time and $a(t)$ is the scale factor. We mostly use the 
conformal time $\eta$ which is related to cosmic time via  $\dd t=a \dd \eta$. We do not specify the form of $V(\phi)$, we only require that it is flat enough 
to sustain inflation for about 60 e-folds. 

We impose  the radiation-Coulomb gauge, $A_0=\partial_iA^i=0$, and   introduce the ``physical"  electric and magnetic fields associated  to the vector potential 
${\bf A}\left(\eta,\,\bfx\right)$
via
\begin{eqnarray}
\label{E-B-Def}
{\bf E} \equiv -\dfrac{1}{a^2} \partial_\eta{\bf A} \,,
\hspace{3cm}
{\bf B} \equiv \dfrac{1}{a^2} {\boldsymbol \nabla} \times {\bf A} \,.
\end{eqnarray}

The dynamics of the system are given by the  Friedmann, Klein-Gordon (KG) and Maxwell equations which are given respectively by  
\begin{align}
\label{Friedmann}
&3 \Mp^2 H^2 = \dfrac{1}{2}\dot{\phi}^2+V +\rho_{\rm em} \,;
\hspace{4cm} \rho_{\rm em} \equiv\dfrac{1}{2}\left(E^2+B^2\right) \,,
\\
\label{KG}
&\ddot{\phi}+3\,H\,\dot{\phi}-\dfrac{\nabla^{2}}{a^2}\phi+V_{,\phi}=J \,;
\hspace{4.1cm}
J \equiv \frac{\alpha}{f} \ 
\boldsymbol{E} \cdot \boldsymbol{B}
\,,
\\
\label{E}
&\dot{\bf E}+2\,H\,{\bf E}-\dfrac{1}{a} \ {\boldsymbol \nabla} \times {\bf B} =-\frac{\alpha}{f}\left( \dot{\phi}\,{\bf B}+{\boldsymbol \nabla}\phi \times {\bf E}\right)\,,
\\
\label{kE}
&{\boldsymbol \nabla} \cdot {\bf E} =-\frac{\alpha}{f} {\boldsymbol \nabla}\phi \cdot {\bf B} \mbox{ ,}
\end{align} 
Moreover, the Bianchi identities read
\begin{eqnarray}
\label{B}
&&\dot{\bf B}+2\,H{\bf B}+\dfrac{1}{a} \ {\boldsymbol \nabla} \times {\bf E} = 0 \,,
\\
\label{kB}
&&{\boldsymbol \nabla} \cdot {\bf B}=0 \,.
\end{eqnarray}
Here $ H \equiv \dot{a}(t)/a(t)$ is the Hubble parameter during inflation and the a dot denotes derivative  with respect to  $t$. 

The basic picture of inflation at the background level in the above setup is as follows. Originally, the gauge field  has no classical background value  so inflation is driven by the inflaton field rolling slowly on top of its potential. Because of the parity violating interaction  \eqref{int} one polarization of the gauge field perturbations 
becomes tachyonic and grow exponentially during inflation while the other polarization is  damped. The tachyonic growth of the gauge field can affect the slow-roll  dynamics in two different ways. First, the electric field energy density associated with the gauge field fluctuations can become significant so the source term $ \rho_{\rm em}$ in the Friedmann equation (\ref{Friedmann}) can not be ignored.  This is the backreactions of the gauge field on the geometry or the Hubble expansion rate. The second effect is the backreaction of the tachyonic gauge field on the slow-roll  dynamics of the inflaton field. Specifically, the source term $J$ in Eq. (\ref{KG}) can become comparable to the driving term $V_\phi$ violating the slow-roll conditions prematurely. Therefore,  in order to make sure that the slow-roll inflation at the background level is not destroyed, we require that both of the above two backreactions to be under control \cite{Barnaby:2010vf,Barnaby:2011vw,Barnaby:2011qe, Linde:2012bt, Gorbar:2021rlt, VargasMoniz:2010upl}.

To study the background evolution, it is more convenient to introduce the slow roll parameters as follows, 
\begin{align}
\label{slowroll}
\epsilon_{_H} \equiv -\dfrac{\dot{H}}{H^2}
\,,
\hspace{2cm}
\epsilon_{_\phi} \equiv \dfrac{\dot{\phi}^2}{2\Mp^2 H^2} \,,
\hspace{2cm}
\epsilon_{_V} \equiv \dfrac{\Mp^2}{2} \left(\dfrac{V_{,\phi}}{V}\right)^2
\,.
\end{align}
These parameters are small in conventional inflationary models and are nearly coincident to a good accuracy when the backreactions of the gauge field are small. 
But in the presence of large electromagnetic fields, these parameters do not coincide in general.  Using Eqs.~\eqref{Friedmann}-\eqref{kB}, it can be shown that
\begin{align}
\label{epsilon}
\epsilon_{_H} = 
\epsilon_{_\phi}+\dfrac{2\rho_{\rm EM} }{3\Mp^2H^2}+\left(\dfrac{{\boldsymbol \nabla}}{aH}\right)\cdot \dfrac{ {\bf E}\times {\bf B}}{6\Mp^2H^2} \,.
\end{align}

In  subsections \ref{Backreaction}, 
we study the slow-roll and backreaction conditions in more details.


\subsection{Production of gauge field fluctuations}
\label{Production}

The equation of motion for the components of the vector field is given by
\begin{eqnarray}
\label{EoM-A}
\Big(\partial_\eta^2-\nabla^2-\frac{\alpha\partial_\eta\phi}{f}{\boldsymbol \nabla}\times \Big){ A_i}(\eta,\bfx)=0 \, .
\end{eqnarray} 
This equation describes the production of the quanta of gauge fields through its coupling to the inflaton field. 

 We decompose the operators  $A_i(\eta,\bfx)$ into the annihilation and creation operators $\hat{a}^\lambda_{\boldsymbol{k}}$ and $\hat{a}^{\lambda \dagger}_{-\boldsymbol{k}}$ as follows 
\begin{eqnarray}
\label{A}
&& {\boldsymbol A}(\eta,\bfx) = \sum_{\lambda = \pm} \ \int \frac{{\rm d}^3k} {\left(2\pi\right)^3} \,  e^{i\boldsymbol{k}.\bfx} ~
\boldsymbol{e}_\lambda(\hat{\boldsymbol{k}}) \left[ A^\lambda(\eta,\boldsymbol{k}) \,\hat{a}^\lambda_{\boldsymbol{k}} +  A^{\lambda *}(\eta,\boldsymbol{k})\, \hat{a}^{\lambda \dagger}_{-\boldsymbol{k}} \right] \,,
\end{eqnarray}
in which $\boldsymbol{e}_\lambda$ are the circular polarization vectors satisfying the following relations
\begin{eqnarray}
\boldsymbol{e}_\lambda(\hat{\boldsymbol{k}}).\boldsymbol{e}_{\lambda'}^*(\hat{\boldsymbol{k}})&=&\delta_{\lambda\lambda'} \,,\\
\boldsymbol{\hat{k}}.\boldsymbol{e}_\lambda(\hat{\boldsymbol{k}}) &=& 0 \,,\\
i\hat{\boldsymbol{k}} \times \boldsymbol{e}_\lambda &=& \lambda \boldsymbol{e}_\lambda \,,
\label{k-cross-e}
\\
\boldsymbol{e}_\lambda(\hat{\boldsymbol{k}}) &=& \boldsymbol{e}^*_{\lambda}(-\hat{\boldsymbol{k}}) \,,
\\
\sum_{\lambda = \pm} e_i^{\lambda}(\hat{\boldsymbol{k}})~e_j^{\lambda *}(\hat{\boldsymbol{k}}) &=& \delta_{ij}-\hat{k}_i \hat{k}_j \,.
\end{eqnarray}
Inserting the decomposition \eqref{A} into  
Eq. \eqref{EoM-A} and assuming $a\left(\eta\right)\simeq -1/(H\,\eta)$ leads to the following equation of motion for the mode functions $A^\lambda(\eta,k)$,
\begin{align}
\label{Al}
\Big(\partial_\eta^2+k^2+\dfrac{2\lambda\xi}{\eta}k\Big)
A^\lambda(\eta,k)=0 \,,
\end{align}
in which $\xi$ is known as the \textit{instability parameter}, defined via the relation
\begin{align}
\label{xi}
\xi  \equiv \dfrac{\alpha \ \dot{\phi}}{2fH} =  \dfrac{\alpha}{f} \Mp \sqrt{\dfrac{\epsilon_{_\phi}}{2}} \,,
\end{align}
where in the second equality we have used the definition of $\epsilon_\phi$ presented in Eq. \eqref{slowroll}. 

It is easy to show that
\begin{align}
\dfrac{\dot{\xi}}{H\xi} \simeq \dfrac{\ddot{\phi}}{H\dot{\phi}}+\epsilon_{_H} \,,
\end{align}
which means that if the slow-roll conditions are satisfied then $\xi$ is nearly constant. 
However, if the deviation from slow-roll becomes noticeable then for two different modes $k_1\ne k_2$, leaving the horizon at different times, the instability parameter $\xi$ takes different values with $\xi_1 \ne \xi_2$. But during the time when a given mode leaves the horizon we may treat $\xi$ to be constant while considering its adiabatic  evolution towards later time during inflation.

The  mode function equation  \eqref{Al} shows that the two polarizations are treated differently through their interactions with the inflaton field.  Without loss of generality, let us suppose $\dot{\phi}>0$ during inflation so $\xi>0$. 
Correspondingly, the positive-helicity mode $A_+(\eta)$ experiences tachyonic
instability for the modes with $k<k_{\rm cr}$ in which
\begin{align}
k_{\rm cr}\equiv \dfrac{2\xi}{|\eta|}\,.
\end{align}
The tachyonic growth of the modes $A_+(k<k_{\rm cr})$  can also be seen from the 
general solution of \eqref{Al}, given by
\begin{eqnarray}
\label{Alambda}
A^\lambda(\eta,k) = \dfrac{e^{\lambda \pi \xi/2}}{\sqrt{2k}} \ {\rm W}_{-i\lambda \xi , \frac{1}{2}}(2ik\eta) \,,
\end{eqnarray}
obtained by imposing the  Bunch-Davies (Minkowski) initial condition for the modes deep inside the horizon\footnote{Note that $e^{-ik\eta} \ \left(2k\eta\right)^{i\xi \lambda} = e^{-i\left(k\eta-\xi\lambda\ln(2k\eta)\right) } \simeq e^{-ik\eta}$. }  
$A_{\lambda}(k\eta \rightarrow -\infty)\simeq e^{-ik\eta}/\sqrt{2k}$  in which ${\rm W}$ is the regular Whittaker function. Consequently, the tachyonic growth of the mode 
$A_+( k < k_{\mathrm{cr}})$ is approximately given by\footnote{ We have used the relation $W_{-\kappa,\mu}(z) \cong \left(\dfrac{z}{\kappa}\right)^{1/4}\kappa^{-\kappa}e^{\kappa}e^{-2\sqrt{\kappa z}}$ for large $|\kappa|$ and when $\Im(\kappa) > 0$.}
\cite{Anber:2009ua}
\begin{align}
\label{amplified_A}
A_+(\eta,k) \cong \dfrac{1}{\sqrt{2k}} \ \left( \dfrac{-k \eta}{2\xi} \right)^{1/4} \exp(\pi \xi-2\sqrt{-2\xi k \eta}) \,,
\hspace{1cm}
(8\xi)^{-1} \lesssim -k \eta \lesssim 2\xi  \,.
\end{align}

The exponential enhancement factor $e^{\pi \xi}$ reflects the nonperturbative nature of the gauge field particle production in the regime $\xi \gtrsim 1$ which we assume throughout. As seen from Eq. \eqref{amplified_A}, the mode function  $A_+$ has a 
real value. This fact displays the classical nature of the produced gauge modes for $k<k_{\rm cr}$ in the sense that 
\begin{align}
\label{classic}
\left[ A_i(\eta,\bfx), \partial_\eta A_j(\eta,\bfx')\right] \approx 0 \,.
\end{align}
This relation indicates the classical evolution of the gauge field fluctuations. 

The tachyonic growth of the gauge field perturbations can backreact on the background geometry and on the evolution of the inflaton field. In the next subsection we deal with this issue in some details. 


\subsection{Backreaction Effects}
\label{Backreaction}

The accumulative  backreactions of tachyonic modes $A_+(k < k_{\rm cr})$ can affect the background evolution. As discussed previously, we have two type of backreactions: the backreaction on the background geometry through the Friedmann equation and the backreaction on the inflaton field in the KG equation. 
Assuming the slow-roll conditions, these two equations take the following forms 
\begin{align}
	\label{bg}
3 \Mp^2 H^2 \simeq V + \rho_{\rm em} \,,
\hspace{2cm}
3H\dot{\phi} \simeq -V_{,\phi} + J \,.
\end{align}
The regime of small backreactions correspond to the situation where both 
$ \rho_{\rm em}$ and $J$ are small. These conditions of small backreactions 
can be parameterized as follows
\begin{align}
\label{Constrant_KG}
R_{J} &\equiv \abs{\dfrac{J}{3H\dot{\phi}}} \ll 1 \,,
\\
\label{Constraint_Friedmann}
\Omega_{\rm em} &\equiv \dfrac{\rho_{\rm em}}{3 \Mp^2 H^2} \ll 1 \,. 
\end{align}
As we shall see later, typically the backreaction on the inflaton dynamics in KG equation becomes important sooner so the condition $R_{J} \ll 1$ is violated earlier. Using the 
solution \eqref{Alambda} in Eq. \eqref{E-B-Def} we can calculate 
 $J \propto \boldsymbol{E} \cdot \boldsymbol{B} $ and $2 \rho_{\rm em} = E^2+B^2$ and look for the effects of backreactions.

In the conventional methods studied in previous works, the backreaction of the amplified gauge quanta on the background  dynamics of $\phi(t)$ and $a(t)$ are taken into account via  \textit{quantum expectations values}. Specifically, considering the modes which experience the tachyonic growth in the regime $\xi \gg 1$ one obtains \cite{Anber:2009ua},
\begin{align}
\label{EdotB}
\langle \boldsymbol{E} \cdot \boldsymbol{B} \rangle &\simeq \dfrac{-1}{2a^4} \int_{}^{} \dfrac{{\rm d}k^3}{(2\pi)^3} k \ \partial_\eta \abs{A_+}^2
\simeq -\dfrac{7!}{2^{21}\pi^2} \left(\dfrac{H}{\xi}\right)^4 \ e^{2\pi\xi} \,,
\\
\label{E2B2}
\dfrac{1}{2}\langle E^2+B^2 \rangle &\simeq \dfrac{1}{2a^4} \int_{}^{} \dfrac{{\rm d}k^3}{(2\pi)^3} \left(k^2\abs{A_+}^2+  \abs{\partial_\eta A_+}^2\right)
\simeq \dfrac{6!}{2^{19}\pi^2} \dfrac{H^4}{\xi^3} \ e^{2\pi\xi} \,,
\end{align}
where $\langle {\cal O} \rangle$ represents the quantum expectation value for the operator ${\cal O}$. The main contribution to the integrals above comes from the scales $k\simeq k_{\rm cr}$. Here, the mean field approximation is assumed in order to construct a homogeneous background from the amplified gauge field fluctuations. The main assumption here is that the accumulative effects of the tachyonic modes generate classical sources $J$ and $\rho_{\rm em}$ which can affect the background dynamics.  It is worth mentioning that  the sign of $\langle \boldsymbol{E} \cdot \boldsymbol{B} \rangle$ is always opposite to the sign of $\dot{\phi}$ so the the tachyonic enhancement of gauge field perturbations 
can actually  prolong the period of inflation. 

Using the  estimations given in Eqs. \eqref{EdotB} and \eqref{E2B2} the conditions of small backreactions from Eqs. (\ref{Constrant_KG})  and \eqref{Constraint_Friedmann}  are translated into 
\begin{align}
\label{cons1}
\xi^{-\frac{3}{2}} \ e^{\pi \xi} &\ll 79 \  \dfrac{\dot{\phi}}{H^2} \,,
\\
\label{cons2}
\xi^{-\frac{3}{2}} \ e^{\pi \xi} &\ll 146 \ \dfrac{\Mp}{H} \,.
\end{align}
Comparing the above two constraints one can check that the former is stronger  than the latter so the backreaction on the KG equation becomes important sooner than the backreaction on the Hubble expansion rate as mentioned before.  

One can use the first condition above to obtain an upper bound on the instability parameter $\xi$.  Imposing the COBE normalization for the power spectrum of curvature perturbation $\zeta \equiv -\frac{H}{\dot{\phi}}\delta \phi$,  
\begin{align}
\label{calP0}
\calP_\zeta^{(0)} \equiv  \left(\dfrac{H}{\dot{\phi}}\right)^2 \left(\dfrac{H}{2\pi}\right)^2 \simeq 2.1 \times 10^{-9} \,,
\end{align}
the  condition (\ref{cons1}) requires $\xi < 4.7\,$.  
Note that the backreaction constraint Eq. (\ref{cons1}) is obtained at the background level. However, as we shall see, the backreaction at the perturbation level (i.e. the effects of gauge field perturbations  on  the CMB-scale curvature perturbations)  puts stronger bound  on $\xi$. 


The instability parameter $\xi$ given in Eq. \eqref{xi} evolves adiabatically during inflation. Here we examine how the evolution of $\xi$ is affected by the tachyonic growth of the gauge field quanta. Using Eq. \eqref{bg}, we can rewrite the instability parameter  in the following form\footnote{Remember that without loss of generality, we assume $\dot{\phi}>0$ during inflation, i.e. $V_{,\phi}<0$ and $\xi>0$. }
\begin{align}
\xi &= 
\xi_V \left(
1+ \langle R_J \rangle
\right)^{-1}
\,;
\hspace{2cm}
\xi_V \equiv \dfrac{\alpha}{f} \Mp \sqrt{\dfrac{\epsilon_{_V}}{2}} \, .
\label{xi3}
\end{align} 
In the absence of strong backreaction when the condition $R_J \ll 1$ is satisfied  
we  have $\xi \simeq \xi_{V}$ which simplifies our calculations. However, as 
gauge modes become tachyonic, the parameter $R_J$ grows like $R_J \propto e^{2\pi \xi}$. Correspondingly, the rapid growth of $\xi$ will affect both the slow-rolling of the inflaton field and the  adiabatic evolution of $\xi$ itself. 
Although a significant backreaction on the inflaton slow-roll dynamics can terminate inflation but  the backreaction on $\xi$ only reduces its rate of change and is not destructive.  In  other words, before the tachyonic instability of the gauge field perturbations  become too significant to destroy the slow-roll evolution of the inflaton field, it first modifies the  evolution of $\xi$ itself,  reducing its growth in such a way that the requirement $R_J \ll 1$ remains valid until close to the end of inflation. We may refer the backreaction on $\xi$ as  the mild backreaction regime because the backreaction conditions  on the background dynamics 
\eqref{Constrant_KG} and \eqref{Constraint_Friedmann} remain valid and $\xi$ will not be larger than ${\cal O}(10)$. 


\subsection{Power spectrum}
\label{Power}

The modification of the scalar power spectrum due to the tachyonic growth of the gauge field perturbations has been studied in \cite{Barnaby:2010vf,Barnaby:2011vw,Barnaby:2011qe, Linde:2012bt}. Here we follow the estimation which has been presented   in App.~B of Ref.~\cite{Linde:2012bt} which is obtained using a semi-analytic approach. 

According to \cite{Anber:2009ua}, the equation of motion for the  inflaton perturbations is given by
\begin{align}
\label{Eom_Linde}
\delta \ddot{\phi}(t,\bfx)+3\tilde{\beta} H \delta\dot{\phi}(t,\bfx)-\dfrac{\nabla^2}{a^2}\delta \phi(t,\bfx)+V_{,\phi \phi}\delta \phi(t,\bfx) = {\cal J}(t,\bfx)\,,
\end{align}
where the effective friction coefficient, $\tilde{\beta}$, and the source term, $\cal J$, are given by
\begin{align}
\tilde{\beta} \equiv 1- 2\pi \xi 
\dfrac{\langle 
	J
	\rangle}{3H\dot{\phi}} =1+ 2\pi \xi \langle R_J \rangle\,,
 \qquad \qquad \qquad  
{\cal J} \equiv J -\langle J \rangle \,.
\end{align}
In our convention where $\dot{\phi}>0$ and $\langle J \rangle<0$, the production of gauge quanta results in an additional friction on the inflaton motion that prolongs the duration of inflation. 

The scalar power spectrum contains two parts:
\begin{align}
\calP_\zeta(k) = \calP_\zeta^{(0)}(k) + \Delta \calP_\zeta(k) \, ,
\end{align}
in which $\calP_\zeta^{(0)}$ is the power spectrum from the scalar vacuum fluctuations defined in Eq. \eqref{calP0} and $\Delta \calP_\zeta(k)$ is the contribution of the source term $J$.  We  estimate  $\Delta \calP_\zeta(k)$ as follows.  Near the horizon crossing, the first term on the left hand side of 
Eq. \eqref{Eom_Linde} cancels the third one. Furthermore, discarding the last term  in the slow-roll regime yields the following estimation for the correction in the  inflaton perturbation induced by the source, 
\begin{align}
\label{phi-J}
\delta\phi_J \approx \dfrac{\cal J}{3\tilde{\beta} H^2} \,.
\end{align}
Using the relation $\zeta \equiv -\frac{H}{\dot{\phi}}\delta \phi$ for  the curvature perturbation on the flat hypersurfaces,   the induced curvature perturbation by the source is given by 
\begin{align}
\langle \zeta_J^2 \rangle \simeq  
\dfrac{\langle {\cal J}^2 \rangle}{9\tilde{\beta}^2 H^2 \dot{\phi}^2} \,.
\end{align}
In fact, the numerator corresponds to the variance $\delta({\bf E}\cdot{\bf B})^2 \equiv \langle ({\bf E}\cdot{\bf B})^2 \rangle - \langle {\bf E}\cdot{\bf B}\rangle^2$. 
A good estimation for the variance has been calculated in App.~B of Ref.~\cite{Linde:2012bt} yielding  $\delta({\bf E}\cdot{\bf B})^2 \simeq \langle {\bf E}\cdot{\bf B}\rangle^2$. Putting all results  together one finds
\begin{align}
\label{DeltaP}
\calP_\zeta(k) \simeq \calP_\zeta^{(0)}(k) +
\left(
\dfrac{\langle R_J \rangle}{\tilde{\beta}}
\right)^2\,.
\end{align}

In the regime of no strong backreaction  where $\tilde{\beta} \simeq 1$, the relative correction to the power spectrum is given by
\begin{align}
\label{DP_estimation}
\dfrac{\Delta \calP_\zeta(k)}{\calP_\zeta^{(0)}} \simeq 4.0 \times 10^{-5} \ 
\calP_\zeta^{(0)} \
\dfrac{e^{4\pi\xi}}{\xi^6} \,,
\end{align}
in which the relations \eqref{EdotB} and \eqref{calP0} have been used. 

For comparison, the fractional  power correction has also been computed by the Green function method in \cite{Barnaby:2010vf,Barnaby:2011vw}, obtaining
\begin{align}
\label{DeltapowerBarnaby}
\dfrac{\Delta \calP_\zeta(k)}{\calP_\zeta^{(0)}}
\cong {\cal P}_\zeta^{(0)} e^{4\pi\xi} \times
\begin{cases}
3 \times 10^{-5} ~ \xi^{-5.4} & 2 \le \xi \le 3 \\
7.5 \times 10^{-5} ~ \xi^{-6} & \xi \gg 1
\end{cases} \,.
\end{align}
As seen, the quick estimation yielding to Eq.  \eqref{DP_estimation} is off from the more extensive analysis yielding to Eq. (\ref{DeltapowerBarnaby}) by a factor less than 2 in  the large $\xi$ limit. From either of the above results we conclude that 
the backreaction on the evolutions of  $\delta \phi$ at the perturbation level 
becomes important earlier (around  $\xi \sim 2.7$) than the backreaction on the background homogeneous equation of $\phi$ obtained 
from Eq. (\ref{cons1}) (around  $\xi \sim 4.7$).  

When the system has reached the regime where  $\tilde{\beta} \gg 1$ then one can approximate $\tilde{\beta} \approx 2\pi \xi  \langle R_J \rangle$ which  from 
Eq. (\ref{DeltaP}) immediately gives
\begin{align}
\Delta \calP_\zeta(k) \simeq \dfrac{1}{\left(2\pi \xi\right)^2} \,.
\end{align}
This equation suggests that curvature perturbations  on small scales, where the backreactions effects become important, are much larger than those on  the CMB-scales. For example,  assuming $\xi \sim 5$ near the end of inflation, the amplitude of the curvature perturbations is estimated to be  $\calP_\zeta \sim {\cal O}(10^{-2})$.

Before ending this Section we comment that in the above analysis the 
backreactions are estimated by integrating over tachyonic  quantum modes  to construct a classical quantity. 
For example,  the quantum expectation value of 
$\langle J \rangle  \propto \langle \boldsymbol{E} \cdot \boldsymbol{B} \rangle$ has been constructed by the contribution of the tachyonic modes to act as a classical source for the background evolution of the inflaton in the KG equation. Similarly, 
the expectation value  $\langle \boldsymbol{E}^2 + \boldsymbol{B}^2 \rangle$ has been calculated to act as a source for the evolution of the background Friedmann equation. This approach looks reasonable and  one expects that the resulting 
estimation for the magnitudes of the backreactions to be reliable.

Our goal in the next Sections is to study this question using the alternative approach of stochastic inflation. One important deviation in our analysis  is that we construct the background electromagnetic fields from the equilibrium states of their coarse-grained field values.
This, in conjunction with the effects of the noises, enable us to provide a new estimation for  the backreactions.   However,  as we shall argue below, the order of the magnitudes of both estimations for the backreaction are consistent with each other.   
 
\section{Stochastic Formalism}
\label{Eom-long}

Here we briefly review the formalism of stochastic inflation and then apply it for our 
setup of axion inflation. 

Stochastic inflation is an effective theory for the evolution of 
long modes on superhorizon scales which are continuously under the influences of the small scale modes which cross the horizon. The effects of these small scale modes  upon horizon crossing can be captured by  Gaussian white noises with the amplitude $H/2 \pi$ \cite{Starobinsky:1986fx,Sasaki:1987gy, Nambu:1988je, Nakao:1988yi, Rey:1986zk, Starobinsky:1994bd}.
To study the dynamic of a field in the stochastic formalism, we split the field and its conjugate momentum into the long IR and the short UV modes. This decomposition is performed via the Heaviside function $\Theta(k-k_{c})$ in which $k_{c}$ is a cutoff scale for IR modes. Using this window function for a generic field $X(t,\bfx)$ and its momentum $\Pi(t,\bfx)$, we have \cite{Sasaki:1987gy, Nambu:1988je, Nakao:1988yi}
\begin{align}
\label{X_decomposition}
X(t,\bfx) &= X_l (t,x)+ \int \dfrac{{\rm d}^3k}{(2\pi)^3} \ e^{i{\boldsymbol k}.\bfx} \ \Theta(k-k_{c})  \ X_{\boldsymbol k}(t) \,,
\\
\label{Pi_decomposition}
\Pi(t,\bfx) &= \Pi_l (t,x)+ \int \dfrac{{\rm d}^3k}{(2\pi)^3} \ e^{i{\boldsymbol k}.\bfx} \ \Theta(k-k_{c})  \ X_{\boldsymbol k}(t) \,,
\end{align}
where $k_{c} \equiv \varepsilon a(t) H$ in which $\varepsilon$ is a small dimensionless parameter. The mode operator $X_{\boldsymbol k}(t)$ contains the corresponding annihilation and creation operators. 

For the electric, magnetic and scalar fields one can use the following 
expansions for the corresponding quantum mode functions, 
\begin{eqnarray}
&&{\boldsymbol E}_{\boldsymbol k}(t) = \sum_{\lambda = \pm} \ \boldsymbol{e}^\lambda(\hat{\boldsymbol{k}}) \left[ E_\lambda(t,k) \,\hat{a}^\lambda_{\boldsymbol{k}} +  E_{\lambda}^{*}(t,k)\, \hat{a}^{\lambda \dagger}_{-\boldsymbol{k}} \right] \,,
\\
&&{\boldsymbol B}_{\boldsymbol k}(t) = \sum_{\lambda = \pm} \ \boldsymbol{e}^\lambda(\hat{\boldsymbol{k}}) \left[ B_\lambda(t,k) \,\hat{a}^\lambda_{\boldsymbol{k}} +  B_{\lambda}^{*}(t,k)\, \hat{a}^{\lambda \dagger}_{-\boldsymbol{k}} \right] \,,
\\
&&\phi_{\boldsymbol k}(t) = \varphi_{\boldsymbol k}(t) \,\hat{b}_{\boldsymbol{k}} +  \varphi^{*}_{\boldsymbol k}(t)\, \hat{b}^{\dagger}_{-\boldsymbol{k}} \, ,
\end{eqnarray}
where the creation and the annihilation operators associated to the 
inflaton and the gauge field  obey the following commutation relation:
\begin{align}
\left[\hat{b}_{\boldsymbol{k}}, \hat{b}_{\boldsymbol{k}'}^\dagger\right] &= (2\pi)^3 \ \delta^{(3)}({\bf k}-{\bf k'}) \,,
\\
\left[\hat{a}^{\lambda}_{\boldsymbol{k}}, \hat{a}^{\lambda' \dagger}_{\boldsymbol{k}'}\right] &= (2\pi)^3 \  \delta^{\lambda \lambda'} \ \delta^{(3)}({\bf k}-{\bf k'}) \,,
\end{align}
and 
\begin{align}
\left[\hat{b}_{\boldsymbol{k}'}, \hat{a}^{\lambda}_{\boldsymbol{k}}\right] = \left[\hat{b}_{\boldsymbol{k}'}, \hat{a}^{\lambda \dagger}_{\boldsymbol{k}} \right] = 0 \,.
\label{commut_ab}
\end{align}

Applying the decompositions presented in Eqs. \eqref{X_decomposition} and \eqref{Pi_decomposition} into the equations of motions \eqref{KG}, \eqref{E} and \eqref{B} give the evolution of the mode functions $\varphi_{\boldsymbol k}$, $E_\lambda$ and $B_\lambda$ as well as the evolution of the long mode parts denoted by $\Phi, {\boldsymbol E}_l$ and ${\boldsymbol B}_l$. We use the definitions in Eq. \eqref{E-B-Def} to obtain the mode functions of $E_\lambda$ and 
$B_\lambda$  from the gauge field mode function $A_\lambda$ given in Eq. 
\eqref{Alambda} .

For the evolution of scalar mode under the influence of the backreactions from the electromagnetic fields we have 
\begin{eqnarray}
\label{KG_phik}
\ddot{\varphi}_{\boldsymbol k} + 3H\dot{\varphi}_{\boldsymbol k}+\Big(\dfrac{k^2}{a^2}+V_{,_{\phi\phi}}
\Big)\varphi_{\boldsymbol k} = \dfrac{\alpha}{f}\big(
{{\boldsymbol E}}_l
.{{\boldsymbol B}}_{\boldsymbol k}+
{{\boldsymbol B}}_l
.{{\boldsymbol E}}_{\boldsymbol k} \big)  \,.
\end{eqnarray}
Note that the electromagnetic fields appear in the source as a combination of a long mode and a {\it linear} short mode. This is in contrast to the analysis of conventional approach
reviewed in the previous section when the combination 
${{\boldsymbol B}}\cdot{{\boldsymbol E}}$ appears as the {\it non-linear} source term  with the understanding that both fields are tachyonic quantum modes. 
This is a key effect which play significant roles in our analysis below.

The solution of Eq. (\ref{KG_phik}) consists of two parts: the homogeneous solution and the particular solution which is due to the source. In App.~\ref{scalar-mode}, we have presented a solution for the mode function $\varphi_{\boldsymbol k}$.

Inserting the decompositions \eqref{X_decomposition} and \eqref{Pi_decomposition} into Eqs. \eqref{KG}, \eqref{E} and \eqref{B}, the equations of motion for the IR modes of the scalar field $\Phi_l$ and its conjugate momentum  $\Pi_l$ as well as the IR parts of the electric field ${\boldsymbol{E}}_{l}$ and the magnetic field  ${\boldsymbol{B}}_{l}$ are obtained as\footnote{We use the following approximations for the IR modes
	\begin{eqnarray*}
	\nabla \times {{\boldsymbol B}}_l \simeq 0
	\hspace{2cm}
	\nabla \Phi \simeq 0
	\hspace{2cm}
	\nabla \times {{\boldsymbol E}}_l \simeq 0
	\hspace{2cm}
	\nabla^2 \Phi \simeq 0 \, .
	\end{eqnarray*} }
\begin{eqnarray}
\label{philong}
&&\dot{\Phi} = \Pi_l + \sigma_{_\phi}
\\
\label{Pilong}
&&\dot{\Pi}_l - \tau_{_\phi} +3H\Pi_l+V_{,\phi} =J_l \, ,
\hspace{2cm} 
J_l=\dfrac{\alpha}{f} \ {{\boldsymbol E}}_l.{{\boldsymbol B}}_l \,,
\\
\label{Elong}
&&\dot{{\boldsymbol E}}_l - {\boldsymbol \sigma}^{_E} + 2 H  {{\boldsymbol E}}_l = 
-\dfrac{\alpha}{f} \Pi_l{{\boldsymbol B}}_l \, ,
\\
\label{Blong}
&&\dot{{\boldsymbol B}}_l - {\boldsymbol \sigma}^{_B} + 2 H  {{\boldsymbol B}}_l
 = 0 \, ,
\end{eqnarray}
in which $(\sigma,\tau)$ are the quantum noises, given by
\begin{eqnarray}
\label{noisephi}
&&\sigma_{_\phi}(t,\bfx) \equiv -\dfrac{{\rm d}k_{c}}{{\rm d}t} \int \dfrac{{\rm d}^3k}{(2\pi)^3} \ \delta(k-k_{c}) \ e^{i{\boldsymbol k}.\bfx} \ \phi_{\boldsymbol k}(t) \,,
\\
\label{noisepi}
&&\tau_{_\phi}(t,\bfx) \equiv -\dfrac{{\rm d}k_{c}}{{\rm d}t} \int \dfrac{{\rm d}^3k}{(2\pi)^3} \ \delta(k-k_{c}) \ e^{i{\boldsymbol k}.\bfx} \ \dot{\phi}_{\boldsymbol k}(t) \,,
\\
\label{noiseE}
&&{\boldsymbol \sigma}^{_E}(t,\bfx) \equiv -\dfrac{{\rm d}k_{c}}{{\rm d}t} \int \dfrac{{\rm d}^3k}{(2\pi)^3} \ \delta(k-k_{c}) \ e^{i{\boldsymbol k}.\bfx} \ {{\boldsymbol E}}_{\boldsymbol k}(t) \,,
\\
\label{noiseB}
&&{\boldsymbol \sigma}^{_B}(t,\bfx) \equiv -\dfrac{{\rm d}k_{c}}{{\rm d}t} \int \dfrac{{\rm d}^3k}{(2\pi)^3} \ \delta(k-k_{c}) \ e^{i{\boldsymbol k}.\bfx} \ {{\boldsymbol B}}_{\boldsymbol k}(t) \,.
\end{eqnarray}

As shown in App.~\ref{noise}, the vectorial stochastic noises of the electric and magnetic fields $\boldsymbol{\sigmaE}(t,\bfx)$ and  $\boldsymbol{\sigmaB}(t,\bfx)$ are aligned along the  $\hat \bfx$-direction. Therefore, the amplitude of the electromagnetic fields in other directions decay according to Eqs. \eqref{Elong} and \eqref{Blong} and only the components along the $\hat{\bfx}$-direction  become relevant. Moreover, it is more convenient to use the $e$-folding number, ${\rm d}N=H{\rm d}t$, as a clock and introduce the following dimensionless stochastic variables
\begin{align}
{\cal B}(N) \equiv \dfrac{\hat{\bfx} \, . \, {\boldsymbol B}_l(N,\bfx)}{\Mp H} \,,
\hspace{2cm}
{\cal E}(N) \equiv \dfrac{\hat{\bfx} \, . \, {\boldsymbol E}_l(N,\bfx)}{\Mp H} \,,
\end{align}
for the magnetic and the electric fields. 

Putting all things together, one can recast Eqs. \eqref{philong}-\eqref{Blong} into the following Stochastic Differential Equations (SDEs):\footnote{We have considered the slow-roll approximation,
	\begin{eqnarray}
	\nonumber
	\dot{\Pi}_l + 3H\Pi_l \simeq 3H\Pi_l \,,
	\hspace{2cm}
	\tau_{_\phi} + 3 H \sigma_{_\phi} \simeq 3 H \sigma_{_\phi} \,.
	\end{eqnarray}
	to simplify \eqref{philong2}.}
\begin{align}
\label{philong2}
&\dd \Phi(N)= \Big(-\dfrac{V_{,\phi}}{3H^2}+
\dfrac{J_l}{3H^2}\Big)\dd N + D_{\phi} \ \dd W(N) \,,
\\
\label{Elong2}
&\dd {\cal E}(N) = -\Big(2+\big(\dfrac{\alpha}{f}\Mp\big)^2\,  \dfrac{{\cal B}^2(N)}{3} \Big) {\cal E}(N) \ \dd N + \dfrac{\alpha}{f}\dfrac{V_{,{\phi}}}{3H^2}{\cal B}(N) \ \dd N + \DE \ \dd W(N) \,,
\\
\label{Blong2}
&\dd {\cal B}(N) = -2 {\cal B}(N) ~\dd N + \DB \ \dd W(N) \,,
\end{align}
where the subscript $l$ for the long modes 
has been removed for convenience while 
the explicit forms of the diffusion terms $\DE$, $\DB$ and $D_{\phi}$ 
are given in Eqs. \eqref{DX} and \eqref{DPhi}. Here, we have defined a Wiener process $W$ associated with a  normalized classical white noise ${\Xi}$ as
\begin{align}
\label{Wiener2}
\dd {W}(N) \, \equiv \, {\Xi}(N) \, \dd N \,,
\end{align}
where
\begin{align}
\langle {\Xi}(N) \rangle = 0 \,,
\hspace{3cm}
\langle \Xi(N_1) \, \Xi(N_2) \rangle =  \,\, \delta(N_1-N_2)\,.
\end{align}

\begin{figure}[t]
	\centering
	\includegraphics[width=0.9\linewidth]{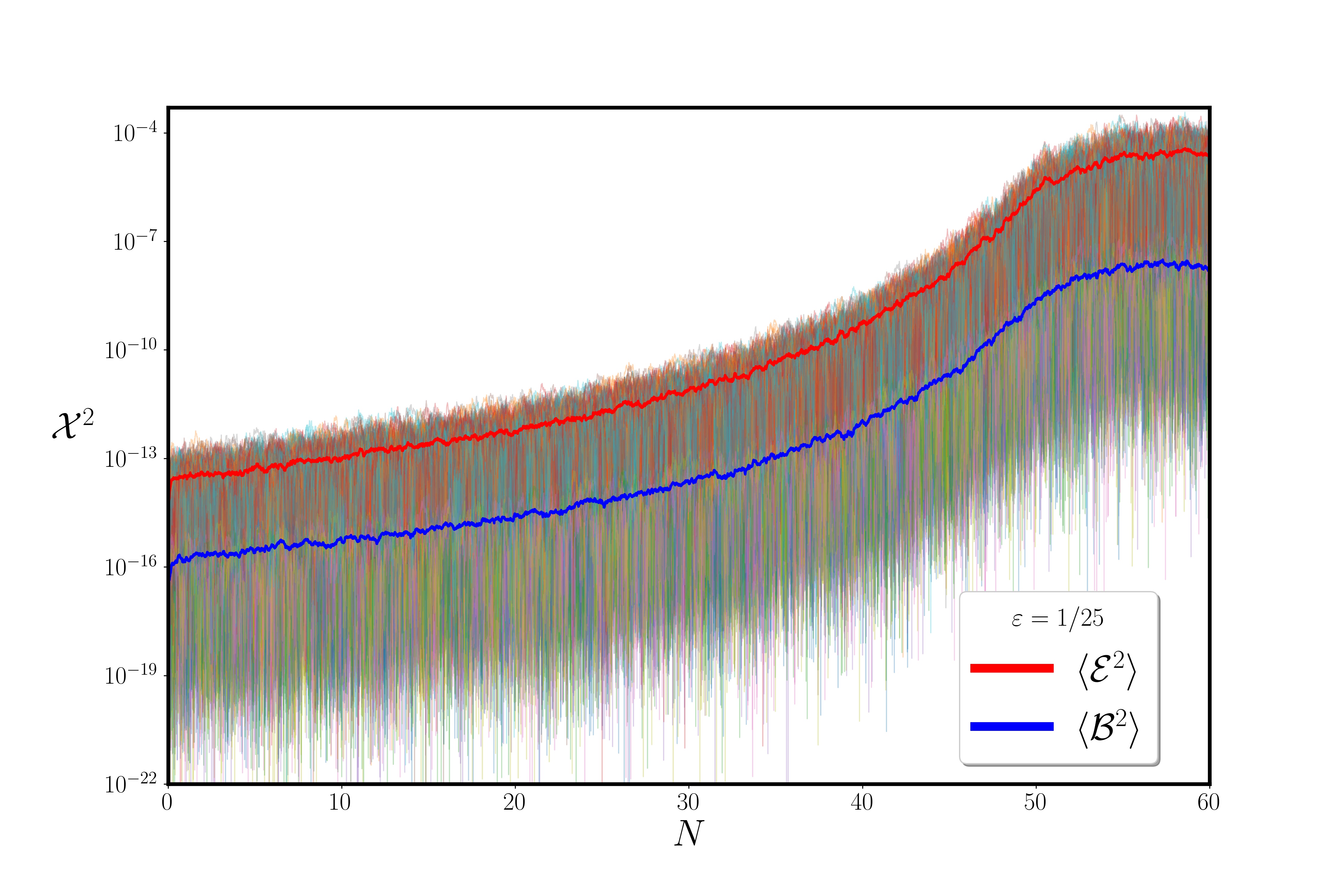}
	\caption{\footnotesize Numerical solution of SDEs  \eqref{Elong2} and \eqref{Blong2} for one hundred realizations. We have considered the scalar potential $V \propto \phi^2$, $\xi_{_{\rm CMB}}=2$, and the Hubble parameter $H = 10^{-6}\Mp$. {The cutoff parameter is chosen via Eq. \eqref{epsilon_lower} to be $\varepsilon \sim \exp(-\pi \xi_{_{\mathrm{CMB}}}/2)$.}  Although the electromagnetic fields show growing behaviours but they have 
 settled down into their local equilibrium states. These growing behaviours are caused from the  adiabatic evolution of $\xi$ (see also Fig.~\ref{fig:Backreaction} for further view) which in turn yields  the growing solution  $ \propto e^{\pi \xi}$ for the electromagnetic fields.	}
	\label{fig:Sto}
\end{figure}

The numerical results for the evolutions of the electric and  magnetic fields are presented in Fig.~\ref{fig:Sto} which show that the electromagnetic fields  settle down  to  local equilibrium states. This is a key property which allows us to construct the background values for the electric and magnetic fields out of the corresponding stochastic variables. The growing behaviour  of the electromagnetic fields seen in Fig.~\ref{fig:Sto} can  be understood through the evolution of $\xi$ (see also Fig.~\ref{fig:Backreaction}) and the growing solution $ \propto e^{\pi \xi}$ for the electromagnetic fields as we will obtain later, see for example  Eqs. 
\eqref{Beq} and \eqref{Eeq}.

In order to describe the time evolution of the electromagnetic fields, one can introduce the Probability Density Function (PDF) of the fields and then employ the Fokker-Planck equation associated with the SDEs \eqref{philong2}-\eqref{Blong2}. 
Intuitively, $f_{\cal X}(x; N) \dd x$ is the probability that the value of ${\cal X}$ falls within the infinitesimal interval $[x, x + \dd x]$ at the given time $N$.
Consider $f_{\cal X}(x; N )$ as the PDF  of the random variable ${\cal X}_N$ which is described by the SDE,
\begin{align}
	\label{SDE}
	\dd {\cal X}_N = \mu({\cal X}_N,N) \ \dd N + D({\cal X}_N,N) \ \dd W_N \, ,
\end{align}
with the drift $\mu$ and diffusion coefficient $D$. The evolution of $f_{\cal X}(x, N )$ is described via the associated Fokker-Planck equation as follows 
\begin{align}
	\label{Fokker}
	\dfrac{\partial f_{\cal X}(x; N)}{\partial N} = -\dfrac{\partial}{\partial x} \Big(\mu(x,N) \ f_{\cal X}(x; N)\Big)+\dfrac{1}{2}\dfrac{\partial^2}{\partial x^2} \Big(D^2(x,N) \ f_{\cal X}(x; N)\Big) \,.
\end{align}

Assuming a constant diffusion coefficient $D({\cal X},N)= D$, there are two simple interesting cases for the drift coefficient as follows:
\begin{itemize}
\item $\boldsymbol{\mu({\cal X},N) = 0}:$\\
	In this case, the SDE \eqref{SDE} describes a Wiener process. The evolution of stochastic variable ${\cal X}$ is often called a Brownian motion and thus the PDF $f_{\cal X}(x; N)$ follows a normal (Gaussian) distribution, denoted by ${\mathbb N}(0,D^2N)$, describing a random walk process with zero mean and variance $D^2N$, at a fixed time $N$:
	\begin{align}
		f_{\cal X}(x; N) = \dfrac{1}{\sqrt{2\pi D^2 N}} \exp(\frac{-x^2}{2D^2 N}) \,.
	\end{align}
In this paper, the parameter space is such that we do not encounter this case, i.e. 
in the current analysis $\mu({\cal X},N) \neq 0$.

\item $\boldsymbol{\mu({\cal X},N) = c -b {\cal X}_N}$ with $\boldsymbol{b>0}:$\\
In this case the SDE \eqref{SDE} represents an Ornstein-Uhlenbeck (OU) process for $\cal X$. During this process, the random force with the amplitude $D$ can balance the frictional drift force $-b {\cal X}$,  washing out the explicit dependence of the mean to the initial conditions ${\cal X}_{\rm ini}$ over time. This process describes the continuous inflow of randomness into the system with the long-term mean $c/b$ and the long term variance $D^2/(2b)$ while the trajectories of ${\cal X}$ evolves around  $c/b$ in the long run. Therefore, the distribution of $\mathcal{X}$ approaches the normal distribution $\mathbb{N}\Big(\frac{c}{b},\frac{D^2}{2b}\Big)$ as $N \rightarrow \infty$.
	
The process tends towards its long-term mean (mean-reverting process), with a greater attraction as the system is further away from the mean. Therefore, the stochastic variable ${\cal Y} \equiv {\mathcal{X}}-\frac{c}{b}$ admits an equilibrium PDF, $\partial f^{\rm eq}_{{\cal Y}}/\partial N=0$, with a bounded variance. The equilibrium solution of Fokker-Planck Eq. \eqref{Fokker} is then given by
	\begin{eqnarray}
		\label{f_Xi_station}
		f^{\rm eq}_{{\cal Y}}(x) = \sqrt{\dfrac{b}{\pi D^2}} ~\exp \bigg({\dfrac{-b \, x^2}{D^2}}\bigg)\,.
	\end{eqnarray}
Equipped with the above PDF, we obtain
	\begin{align}
		\label{XEq}
		\langle {\cal X} \rangle_{\rm eq} = \dfrac{c}{b} \,,
		\hspace{1cm}
		\langle {\cal X}^2 \rangle_{\rm eq} = \left(\dfrac{c}{b}\right)^2+\dfrac{D^2}{2b} \,,
		\hspace{1cm}
		\delta {\cal X}_{\rm eq}^2 &=\dfrac{D^2}{2b} \,,
	\end{align}
in which $\delta {\cal X}^2 \equiv \langle {\cal X}^2 \rangle_{\rm eq} - \langle {\cal X} \rangle_{\rm eq}^2$ is the variance. 

Although the equilibrium time $N_{\rm eq}$ when ${\cal X}$ reaches to the stationary value ${\cal X}_{\rm eq}$ is formally 
infinite, but for our practical purpose we estimate $N_{\rm eq}$ as the time when $\left| \langle \mathcal{X}^2(N_{\rm eq}) \rangle -\langle \mathcal{X}^2 \rangle_{\rm eq} \right|/\langle \mathcal{X}^2 \rangle_{\rm eq}$ drops to a small value say $e^{-q}$ for $q \gtrsim 4$. A good estimation is obtained to be $N_{\rm eq} \simeq q/2b$. Therefore,  $b$ characterizes the speed at which the trajectories will regroup around the mean $c/b$.

When $c$ is a positive constant  the SDE \eqref{SDE} describes an OU process which is known as the Vasicek equation \cite{vasicek1977equilibrium} while for $c=0$ the SDE \eqref{SDE} is called the Langevin equation.
\end{itemize}

Looking at Eqs. \eqref{Elong2} and \eqref{Blong2}  it is clear that the evolution of the magnetic (electric) field is govern by the Langevin (Vasicek) equation. In subsections \ref{sec:B} and \ref{sec:E}, we solve SDEs \eqref{Blong2} and \eqref{Elong2} using the method of Ito calculus. Since the magnetic field is decoupled from the other two fields, first we solve for the magnetic field. 


\subsection{Magnetic field}
\label{sec:B}

In this section, we study the evolution of the long modes of magnetic field  described by Eq. \eqref{Blong2}. As mentioned before, the SDE \eqref{Blong2} describes an OU process in which the frictional drift force $-2{\cal B}$ is balanced by the random force $\DB \Xi$  where $\DB$ is given in Eq. \eqref{DX}. 
Therefore, the distribution of $\cal B$ approaches the normal distribution $\mathbb{N}\big(0,\DB^2/4\big)$ as $N \rightarrow \infty$. With no classical initial condition, the general solution of \eqref{Blong2} is given by
\begin{align}
\label{Bsolint}
{\cal B}(N) = \DB e^{-2N} \int_{0}^{N} e^{2s} \ \dd W(s) \, .
\end{align}
Correspondingly, the average quantities 
$\langle {\cal B}(N)\rangle$ and $\langle {\cal B}^2(N)\rangle$ (after averaging over 
a large number  of simulations) is obtained to be 
\begin{align}
\label{Bsol}
\langle {\cal B}(N)\rangle = 0 \,,
\hspace{2cm}
\langle {\cal B}^2(N)\rangle = \dfrac{\DB^2}{4}\left(1-e^{-4N}\right) \,,
\end{align}
where we have used the following properties of the stochastic integrals \cite{evans2013introduction}, 
\begin{eqnarray}
\label{Ave_StoIntegrals_1}
\Big \langle 
\int_0^{N'} G(N) \, \dd W\int_0^{N'} F(N) \, \dd W
\Big \rangle 
&=& 
\int_0^{N'} G(N)F(N) \, \dd N \,,
\\
\label{Ave_StoIntegrals_2}
\Big \langle
\int_0^{N'} G(N) \, \dd W
\Big \rangle  
&=& 0 \,,
\end{eqnarray}
for the general functions $F$ and $G$. 

As mentioned before, the solution \eqref{Bsolint} admits an equilibrium state which is obtained as $N \rightarrow \infty$. For our practical purpose, we consider 
$N_{\rm eq}^B \lesssim {\cal O}(1)$ as the $e$-folding number of the magnetic field to settle down to its stationary state,
\begin{align}
\langle {\cal B} \rangle_{\rm eq} =0 \,,
\hspace{2cm}
\delta {\cal B}^2 = \langle {\cal B}^2 \rangle_{\rm eq} = \dfrac{\DB^2}{4} \,.
\end{align}
The above solutions are consistent with the stationary solutions \eqref{XEq} obtaining via Fokker-Planck equation when ${\cal X}={\cal B}$, $c=0$, $b=2$ and $D=\DB$. We  denote the amplitude of the magnetic field in the equilibrium state as ${\cal B}_{\rm eq}$, defined as follows
\begin{align}
\label{Beq}
{\cal B}_{\rm eq} \equiv 
\sqrt{\delta {\cal B}^2}= \dfrac{\DB}{2} =  \dfrac{H}{4\Mp} \, \sqrt{\dfrac{\sinh(2\pi\xi)}{3\pi^3\xi}} \, \varepsilon^2 
\end{align}

Since the magnetic field drops to its stationary state very quickly we can replace 
$\cal B$ by its equilibrium value \eqref{Beq} when solving the stochastic differential equations  \eqref{philong2} and \eqref{Elong2} for $\Phi$ and  $\cal E$.

\subsection{Electric field}
\label{sec:E}

Using the solution  \eqref{Beq}, the equation of motion for the  IR modes of electric field \eqref{Elong2} can be rewritten as 
\begin{align}
\label{Elong3}
\dd {\cal E}(N) = \big(c-b \ {\cal E}(N) \big) \ \dd N + \DE \ \dd W(N) \,,
\end{align}
in which
\begin{align}
b &\equiv 2+\big(\dfrac{\alpha}{f} \Mp\big)^2  \dfrac{{\cal B}_{\rm eq}^2}{3} \,,
\hspace{2cm}
c \equiv
2 \xi_V{\cal B}_{\rm eq}  \,,
\end{align}
and  $\xi_V$ is defined in Eq. \eqref{xi3}. The stochastic differential equation \eqref{Elong3} has the form of a  
Vasicek SDE  \cite{vasicek1977equilibrium}. 

In the absence of any classical (initial) electric field, the general solution of 
Eq. \eqref{Elong3} is given by
\begin{align}
{\cal E}(N)= \dfrac{c}{b}\left(1-e^{-b N}\right)+\DE e^{-bN} \int_{0}^{N} e^{bs} \ \dd W(s) \,.
\end{align}
Correspondingly, we  obtain
\begin{align}
\langle {\cal E}(N)\rangle &= \dfrac{c}{b}\left(1-e^{-b N}\right) \,,
\\
\langle {\cal E}^2(N)\rangle &=\dfrac{c^2}{b^2}\left(1-e^{-b N}\right)^2+ \dfrac{\DE^2}{2b}\left(1-e^{-2bN}\right) \,,
\\
\delta {\cal E}^2 &= \langle {\cal E}^2(N)\rangle - \langle {\cal E}(N)\rangle^2 = \dfrac{\DE^2}{2b}\left(1-e^{-2bN}\right) \,.
\end{align}
It's worth mentioning that the equilibrium solutions \eqref{XEq} are consistent with the above results as $N \rightarrow \infty$. In fact, after the $e$-fold $N=N_{\rm eq}^E \lesssim {\cal O}(1)$ the electric field admits a stationery state to a high accuracy. We define the magnitude of the electric field at the equilibrium state as
\begin{align}
\label{Eeq}
{\cal E}_{\rm eq} &\equiv 
\sqrt{\delta {\cal E}^2}= \dfrac{\DE}{2} 
= \dfrac{H}{2\Mp} \, \sqrt{\dfrac{\xi \sinh(2\pi\xi)}{3\pi^3}} \,   \varepsilon^2 
  \abs{\ln\varepsilon} \, .
\end{align}

In conclusion,  we have found that both the electric and magnetic fields settle down to their stationary states very quickly after inflation starts. Therefore, it is justified 
to use Eq. \eqref{Beq} and \eqref{Eeq} for the ``background" values of the 
electromagnetic  fields  in the equation of motion of scalar field 
\eqref{philong2} which is studied in  next Section.

\section{Backreactions and Parameter Space}
\label{backreaction-sec}

We are interested in the small backreactions regime where the tachyonic instability of gauge fields do not have significant effects on the evolution of the inflaton and the background geometry. In the absence of stochastic noises, these requirements lead to the constraints \eqref{cons1} and \eqref{cons2} which impose the upper bound  $\xi < 4.7$ in order to have a long period of slow-roll inflation. In this section we revisit the question of backreactions in our approach based on the formalism of stochastic inflation.  

In the previous section we have shown that the electromagnetic fields settle down to their local equilibrium states. 
Therefore, it is justified  to use their equilibrium values as the source terms for the Friedmann and the KG equations.  Specifically, the backreaction conditions \eqref{Constrant_KG} and \eqref{Constraint_Friedmann}  can be written in terms of the mean values ${\cal B}_{\rm eq}$ and ${\cal E}_{\rm eq}$, defined in \eqref{Beq} and \eqref{Eeq}, as follows 
\begin{align}
	\label{Constrant_KG_sto}
	R_{\rm eq} &\equiv \Big( \dfrac{\alpha }{f}\Mp\Big)^2 \dfrac{{\cal E}_{\rm eq}  {\cal B}_{\rm eq}}{6\xi} \ll 1 \,,
	\\
	\label{Constraint_Friedmann_sto}
	\Omega_{\rm eq} &\equiv \dfrac{1}{6} ({\cal E}_{\rm eq}^2+{\cal B}_{\rm eq}^2) \ll 1 \, .
\end{align}
Using the specific values of ${\cal B}_{\rm eq}$ and ${\cal E}_{\rm eq}$ given in Eqs. \eqref{Beq} and \eqref{Eeq} we obtain 
\begin{align}
\label{cons1_sto}
\xi^{\frac{1}{2}} \ e^{\pi \xi} \  \varepsilon^2 \sqrt{\abs{\ln\varepsilon}} &\ll 47 \  \dfrac{\dot{\phi}}{H^2} \,,
\\
\label{cons2_sto}
\xi^{\frac{1}{2}} \ e^{\pi \xi} \  \varepsilon^2 \sqrt{\abs{\ln\varepsilon}} &\ll 67 \ \dfrac{\Mp}{H} \,.
\end{align}
These conditions should be compared with those obtained in Eqs. (\ref{cons1}) 
and (\ref{cons2}) using the quantum expectation values over the tachyonic modes 
in  Eqs. \eqref{EdotB} and \eqref{E2B2}. The exponential factor $e^{\pi \xi}$ is the same in both sets of formulas. However, in Eqs. (\ref{cons1_sto}) 
and (\ref{cons2_sto}) we also have the contribution from the stochastic parameter 
$\varepsilon$. 
More precisely our estimation of backreactions in  Eqs. (\ref{cons1_sto}) 
and (\ref{cons2_sto}) is different from those in  Eqs. (\ref{cons1})  and (\ref{cons2}) by the combination 
$\xi^2 \, \varepsilon^2 \sqrt{\abs{\ln\varepsilon}}  $. 
We will discuss more about the physical meanings of the parameter 
$\varepsilon$ in the next subsection. 

As before, the constraint (\ref{cons2_sto}) becomes trivial in the limit of slow-roll inflation  because $\dot{\phi} \ll \Mp H$. On the other hand, imposing the COBE normalization  for the scalar power spectrum \eqref{calP0} the  constraint (\ref{cons1_sto}) leads to $\xi^{\frac{1}{2}} \ e^{\pi \xi} \  \varepsilon^2 \sqrt{\abs{\ln\varepsilon}} \ll 1.6 \times 10^5 \,$. As a result, we see that the backreaction condition is now controlled by  two parameters $(\varepsilon, \xi)$.
Hence a relevant question is what is the bound on  $\varepsilon$? In the next subsection, we investigate this question.

\subsection{Cutoff parameter $\varepsilon$}
\label{varep}
As we have seen in the previous analysis, the cutoff parameter $\varepsilon$ plays important roles in estimating the backreaction effects. Specifically, for smaller values of $\varepsilon$ the bound on $\xi$ is relaxed so higher values of $\xi$ are allowed. 
As $\xi$ is the important parameter of the setup which controls the backreactions on the dynamics of the background and the level of induced curvature perturbations and non-Gaussianities, it is important to study the effects of $\varepsilon$ in more details. 

\begin{figure}[t!]
	\centering
	\includegraphics[width=0.8\linewidth]{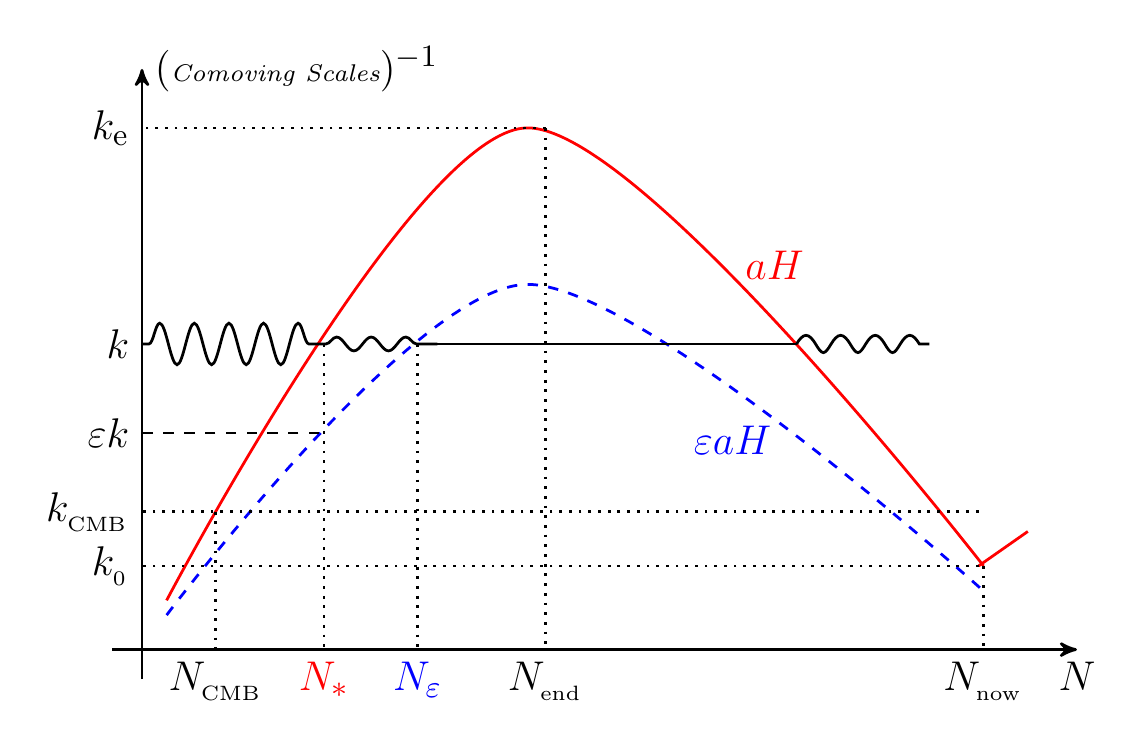}
	\caption{\footnotesize Interpretation of long and short modes in terms of 
	the cutoff parameter $\varepsilon$. When constructing a coarse grained field for the long mode perturbations, all modes in the range  $k > \varepsilon a H$ are integrated out. These modes contribute to the evolution of the coarse grained field through the stochastic noises. During the interval $N_*$ and  $N_\varepsilon$, the mode $k$ becomes superhorizon but can not contribute to the coarse grained field because the mode has not become classical.	}
	\label{fig:ep}
\end{figure}

In the stochastic formalism, the superhorizon  coarse grained field is treated as the background field. In Eq. \eqref{X_decomposition}, we have used the Heaviside function to perform the long and short decomposition so the  coarse grained field  $X_l (t, \bfx)$ is obtained to be
\begin{align}
\label{Xl}
X_l (t,\bfx) &= \int_{0}^{\varepsilon a H} \dfrac{{\rm d}^3k}{(2\pi)^3} \ e^{i{\boldsymbol k}.\bfx}  \ X_{\boldsymbol k}(t) \,.
\end{align}
The coarse grained field contains only  modes with the wave number $k < \varepsilon aH$. With $\varepsilon \ll 1$  the corresponding wavelengths are much longer than the horizon scale $L \sim (aH)^{-1}$.  The coarse grained field is assumed to be a classical field, and since the horizon scale $L$ becomes shorter and shorter, more and more of of subhorizon modes contribute to the coarse grained field and become classical.   By classical we mean that 
the commutator of the field and its conjugate momentum  approaches to zero. For a wide class of models this commutator is proportional to $\varepsilon^m$ with $m>0$. Hence, by choosing $\varepsilon$ small enough the commutator goes to zero and the assumption of classical limit is justified.

Physically speaking, the cutoff parameter $\varepsilon$ represents the  
scale dependency of the electromagnetic fields in our model. To see this, from
Eq. \eqref{BE_lambda} we find that  $B_\lambda \propto \varepsilon^2$ and $E_\lambda \propto \varepsilon^2 \ln\varepsilon$ so the electromagnetic energy density decays as $\varepsilon^4$. This is in line with the fact that the electromagnetic energy density in inflation is diluted as 
$k^{-4}$ at the horizon exit. Another example is the behaviour of the electromagnetic fields in $I^2F^2$ models with $I \propto a^{-n}$ \cite{Talebian:2020drj}. In this case, the amplitude of the diffusion terms or equivalently the power spectrum corresponding to the electric (magnetic) field is given by ${\cal P}_{E} \propto \varepsilon^{2-n}$ ~ (${\cal P}_{B} \propto \varepsilon^{3-n}$). Correspondingly  a scale invariant spectrum for the electric (magnetic) field is obtained  when $n=2$ ($n=3$) as expected. 
Therefore we conclude that the cut off parameter $\varepsilon$ controls  the scale dependency of the system in such a way that $\varepsilon \propto k^{-1}$.

A schematic view of the effects of $\varepsilon$ is presented in Fig.~\ref{fig:ep}. 
Consider a mode $k$ which exits the horizon  $(aH)^{-1}$ at $N_*$ while exiting  the smoothing patch $(\varepsilon aH)^{-1}$ at $N_\varepsilon$. During the interval between $N_*$ and  $N_\varepsilon$ the mode is superhorizon but still it 
 can not contribute to the coarse grained field. In other words, the mode does not become classical at $N_*$ and it  still retains its quantum behaviour  till $N_\varepsilon$.

To estimate the magnitude of $\varepsilon$, let us consider the noises associated to 
the scalar field perturbation given in Appendix \ref{scalar-noises}. From the combination of Eqs. (\ref{cor1})-(\ref{cor3}) we conclude that $\varepsilon$ should satisfy the following constraint $\exp(\frac{-3 H^2}{2 m^2}) \lesssim  \varepsilon \lesssim \frac{m}{H}$ in order for the system to be classical \cite{Sasaki:1987gy, Nambu:1988je, Nakao:1988yi}. On the other hands, from the properties of the electromagnetic noises 
in Appendix \ref{EM-noises}, from the combination  of 
Eqs.  (\ref{Cor1})-(\ref{Cor5}) we require  that 
$ \exp(-\pi \xi/2) \lesssim  \varepsilon \ll   1$ in order for the system to reach the classical limit.  Since typically  $H/m \gg \xi$ during slow-roll inflation 
we conclude that  $\varepsilon$ falls in the following range 
in order for the system to be treated as classical:
\ba
\label{bound-epsilon}
\exp(\frac{-\pi \xi}{2}) \lesssim  \varepsilon  \lesssim   \frac{m}{H} \, .
\ea 
As $\xi$ increases during inflation,  one may consider the smallest value of  $\xi$ at the CMB scale, $\xi_{_{\mathrm{CMB}}}$, such that  the backreactions are under control throughout inflation and  
\begin{align}
\exp(\frac{-\pi \xi_{_{\rm CMB}}}{2}) \lesssim  \varepsilon  \lesssim   \frac{m}{H} \,.
\label{epsilon_lower}
\end{align}
As we shall see from the corrections in power spectrum induced by gauge fields perturbations and the constraints on CMB, the parameter $\varepsilon$ is typically at the order $\varepsilon \lesssim {\cal O}(0.1)$. We will see that this range is also acceptable when we consider the PBH bounds on the power spectrum of the curvature perturbation.

Let us define $\xi_{\rm c}$ as the maximum value of the instability parameter 
where the backreactions on the background Hubble expansion rate in Friedmann equation and the inflaton dynamics in KG equations are negligible.  For examples, by choosing $\varepsilon=\{1/7,1/15,1/25\}$, we have  $\xi_{\rm c}$= $\{4.7, 5.1, 5.4\}$ respectively without encountering the backreactions at the background level. 
This shows  that the  stochastic formalism with a small enough parameter $\varepsilon$ relaxes the upper bound on the  instability parameter $\xi$. 
However, it should be noted that  there is a lower bound on  $\varepsilon$  given by Eq. \eqref{epsilon_lower}. 

\begin{figure}[t]
	\centering
	\includegraphics[width=0.6\linewidth]{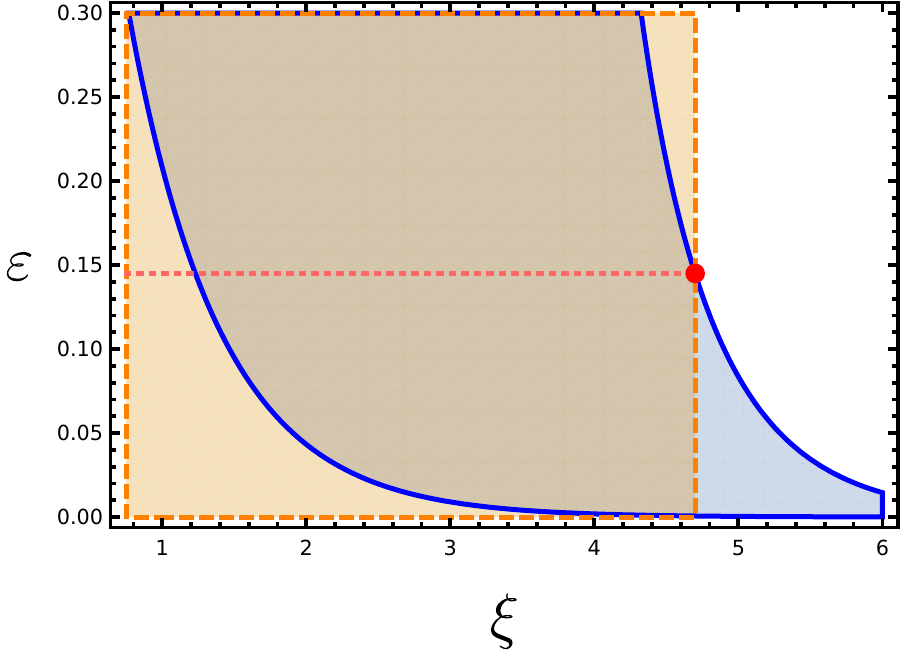}
	\caption{\footnotesize The allowed regions with allowed (i.e. small enough) backreactions  in the stochastic approach (blue) and conventional approach (orange),   see e.g. \cite{Peloso:2016gqs}. The two regions intersect at the red point $(4.7,0.14)$. The right blue boundary curve represents the upper bound on $\xi$, i.e. $\xi_{\rm c}$. Reducing  the cut off parameter $\varepsilon$, the upper bound on $\xi_{\rm c}$ is relaxed in the stochastic approach as seen by the extension of the blue region above. { The left blue boundary curve  represents the lower bound \eqref{epsilon_lower} for $\varepsilon$. For example, for $\xi_{_{\rm CMB}}=\{ 2,3.5 \} $, the cutoff parameter must be chosen so that $\varepsilon \gtrsim \{ 0.04 , 0.004\}$ respectively. However these two values are excluded by PBH constrains on the curvature power spectrum as discussed in Sec.~\ref{sec:PBH}.}}
	\label{fig:ep-xi}
\end{figure}

The allowed parameter space for small backreactions at the background level (i.e. in the absence of perturbations)  are shown in 
Fig.~\ref{fig:ep-xi}. The blue area represents the regions of the parameter space where  the backreactions of the  gauge quanta  on the Hubble expansion rate  and on  inflaton field can be neglected in the stochastic approach. The orange region shows the same  backreaction constraints  if  the conventional methods, such as 
in \cite{Peloso:2016gqs},  are used. 
Our estimation for the backreactions and the allowed range of $\xi$
is qualitatively consistent with the results obtained in conventional approach but the stochastic effects modify the allowed range of  $\xi$ to some extent.  
Further comparisons between the two methods of estimating the backreactions will be given when studying the scalar perturbations in next Section.

\subsection{Instability parameter $\xi$ }
The instability parameter $\xi$ evolves adiabatically during inflation. 
In the presence of the stochastic noises, upon averaging over Eq. \eqref{xi}, we obtain 
\begin{align}
\xi 
&= \dfrac{\alpha}{2f} \left\langle \dfrac{\dd \Phi}{\dd N} \right\rangle
\simeq \xi_V \left(
1+ R_{\rm eq}\right)^{-1} \,,
\label{xi4}
\end{align}
where $\xi_V$ is given by Eq. \eqref{xi3} and we have used Eq. \eqref{philong2} and the stochastic integral \eqref{Ave_StoIntegrals_2}. The behaviours of $\xi$  in 
 Eq.  \eqref{xi4} is similar to Eq. \eqref{xi3}. The only difference is that we  use  $R_{\rm eq}$ instead of  $R_J$ and the upper bound on the allowed range of 
 $\xi$ is somewhat increased in the presence of the  stochastic noises. This is shown in  Fig.~\ref{fig:ep-xi} where  the regime  of  small backreaction  is extended to somewhat larger value of $\xi$ as one allows for smaller values of $\varepsilon$.
 
\begin{figure}[th!]
	\centering
	\includegraphics[width=.7\linewidth]{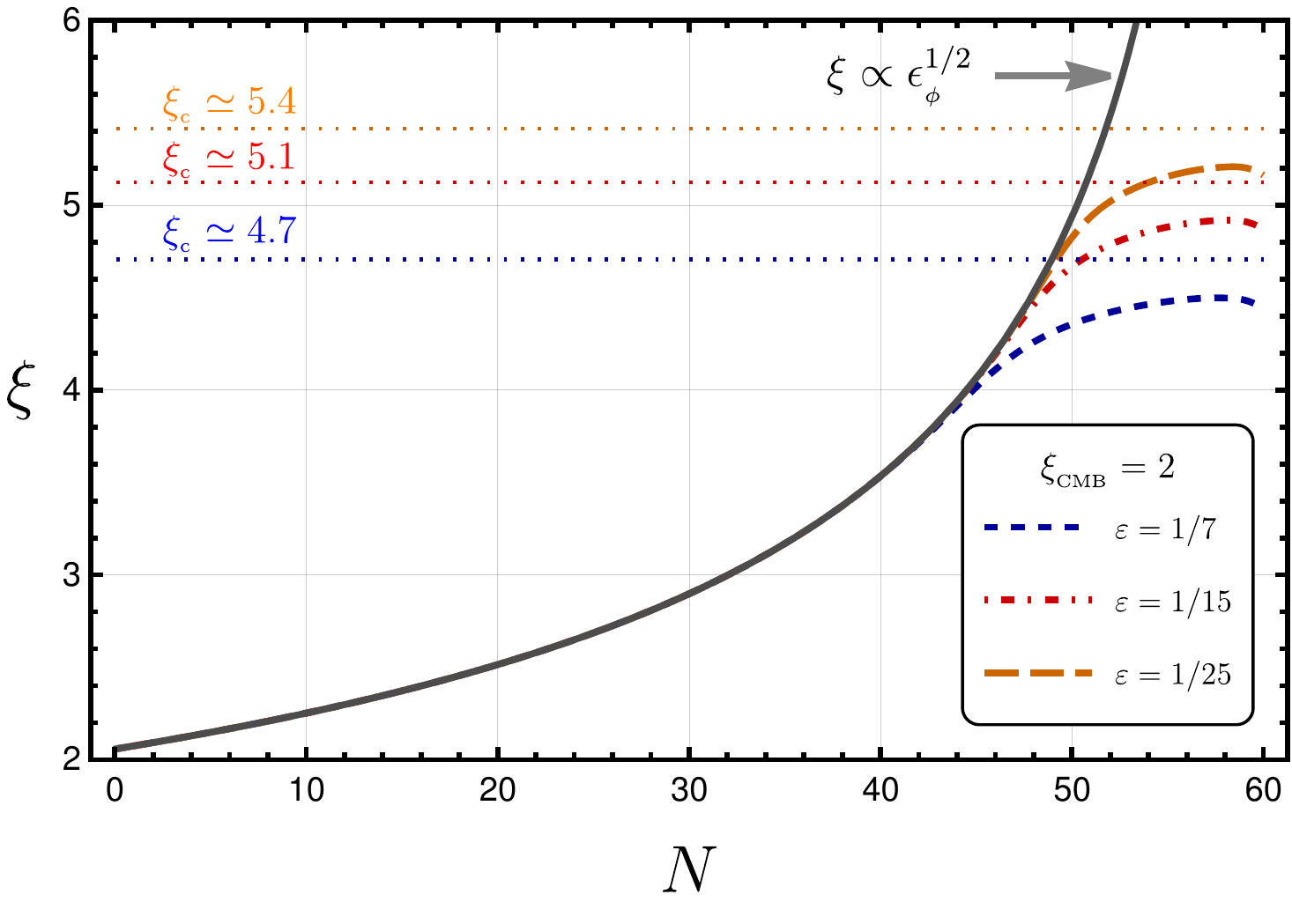}
	\includegraphics[width=.7\linewidth]{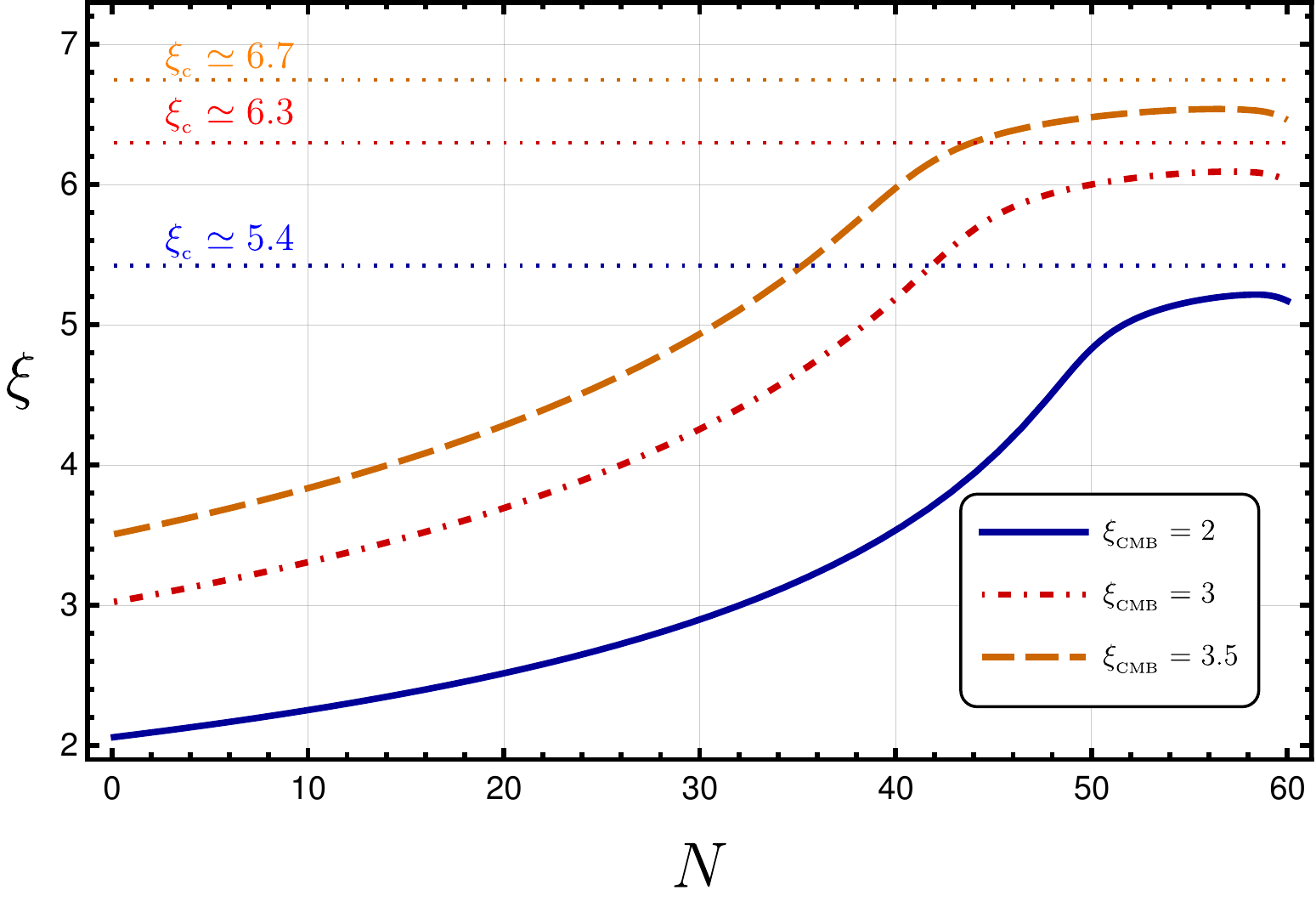}
	\caption{\footnotesize \textbf{Top:} The evolution of the instability parameter in term of $e$-folding number for {$\xi_{_{\rm CMB}}=2$ and quadratic potential when the various cutoff parameter $\varepsilon=1/7$ (blue), ~1/15   (red dash-dotted) and 1/25   (orange long-dashed)} are considered. 
	The plot corresponding to $\varepsilon=1/7$ in our approach with 
 the stochastic noise coincides with the one in the absence of the noises. The horizontal dotted lines represent $\xi_{\rm c}$, the maximum value of $\xi$ which satisfies the constraint $R_{\rm eq} < 1$. {Ignoring the effects of gauge quanta on the evolution of $\xi$ itself is shown by the tick gray solid line. Following this line, the final value of the instability parameter at the end of inflation 
 is simply given by $\xi \sim \alpha/f$  when the first slow-roll parameter is at the order of one.} \textbf{Bottom:} The evolution of $\xi$ for different initial values $\xi_{_{\rm CMB}}$. The parameter $\varepsilon$ for each case is fixed by the lower bound in Eq.  \eqref{epsilon_lower}. For all cases, the final values of $\xi$ is always below the corresponding values of $\xi_{\rm c}$. This means that the system does not experience significant backreactions throughout inflation. }
	\label{fig:Backreaction}
\end{figure}
 
There are two comments in order.  First,  during  inflation $\xi$ grows like $\epsilon_{_\phi}^{1/2}$ while $R_{\rm eq}$ grows like $e^{2\pi \xi}$. Hence, the growth of $\xi$ affects the adiabatic evolution of $\xi$ itself in such a way that around the time of end of inflation $\xi$ approaches  a nearly constant value. Second, the evolution of $\xi$ does not lead to strong backreaction: starting with any values for $\xi_{_{\rm CMB}}$ in the allowed (blue) region of Fig.~\ref{fig:ep-xi}, the system remains in the regime of small backreaction throughout inflation. In other words, the final value of $\xi$ at the end of inflation 
 is always below the maximum value $\xi_{\rm c}$.  These two points are illustrated in Fig.~\ref{fig:Backreaction} where  the evolution of the instability parameter $\xi$  during inflation have been plotted. With this plot, we have also compared  Eq. \eqref{xi3} with the stochastic approach \eqref{xi4}. 
We take the example of large field model $V(\phi)  \propto \phi^2$  with 
$N_{\rm tot}=60$  e-fold of inflation. For the case $\varepsilon=1/7$ the plot of $\xi$ vs. $N$ coincides with that obtained in 
the conventional approach studied in Section \ref{Ratra-Model}.  As seen, for the smaller value of cutoff parameter $\varepsilon$, the backreaction effects on the evolution of $\xi$ starts later so  $\xi$ has enough time to grow to higher value. If one naively ignores the backreactions on $\xi$  itself then inflation is terminated well before $N_{\rm tot}=60$.

\section{Scalar Power Spectrum}
\label{Subsec.Phi}

In this section we study the curvature perturbation power spectrum. In addition to 
the usual contribution from the inflaton perturbations, there is an additional contribution in curvature perturbations induced from the tachyonic  gauge field perturbations. After the electromagnetic fields are settled down to their equilibrium state given by Eqs. \eqref{Beq} and \eqref{Eeq}, the Langevin equation \eqref{philong2} for the long mode of scalar field perturbations takes the following form  
\begin{align}
\label{philong3}
\dd \Phi(N)= \dfrac{\sqrt{2\epsilon_{_V}}\Mp}{1+R_{\rm eq}} \, \dd N + \dfrac{D_{\phi}}{1+R_{\rm eq}} \ \dd W(N) \,,
\end{align}
where the diffusion coefficient $D_{\phi}$ is given by \eqref{DPhi}.

We employ the  stochastic $\delta N$ formalism \cite{Fujita:2013cna,Fujita:2014tja, Vennin:2015hra} to calculate the curvature perturbation 
power spectrum. In $\delta N$  formalism 
\cite{Abolhasani:2019cqw, Lyth:2005fi,Lyth:2004gb,Sasaki:1998ug,Sasaki:1995aw,Starobinsky:1986fxa,Starobinsky:1982ee, Talebian-Ashkezari:2016llx,Talebian-Ashkezari:2018cax},  starting with a flat initial hypersurface, the curvature perturbation is given by  the difference in background number of e-folds between this flat hypersurface and the final hypersurface of constant energy density via 
\begin{align}
	\zeta(\bfx) = N(t, {\bf x})-\bar{N}(t) \equiv \delta N \,,
\end{align}
where $\bar{N}(t)\equiv \ln \left( \frac{a(t)}{a(t_i)} \right)$ is the unperturbed amount of expansion. 

In the stochastic approach the amount of expansion between these two slices is a stochastic quantity which we denote  by ${\cal N}$. Define $\Phi^*(k)$ as the mean value of the coarse grained field when the  given wavenumber $k$ crosses the Hubble radius.  Let us also denote by ${\cal N}(k)$ the number of $e$-folds realized between $\Phi_*(k)$ and $\Phi_{\rm end}$ when inflation ends 
with the variance 
\begin{align}
	\delta {\cal N}^2(k) \equiv \left\langle {\cal N}^2(k)  \right\rangle - \left\langle {\cal N}(k)  \right\rangle^2 \,.
\end{align}
Then using the stochastic $\delta N$ formalism, the 
curvature perturbation power spectrum is given by \cite{Fujita:2013cna,Fujita:2014tja, Vennin:2015hra}
\begin{align}
	\label{Power2}
	{\cal P}_\zeta(k) = {\cal P}_{\delta {\cal N}}(k) =
	\left.
	\dfrac{\dd \left\langle \delta{\cal N}^2  \right\rangle}{\dd\left\langle {\cal N}  \right\rangle}
	\right\vert_{\left\langle\mathcal{N}\right\rangle=\ln\left(\frac{k_{\rm end} } {k}\right) } \, .
\end{align}

To calculate $\left\langle {\cal N}  \right\rangle$ and $\left\langle \delta{\cal N}^2  \right\rangle$, we write  the  Langevin equation \eqref{philong3} in the following form 
\begin{align}
	\label{Langevin}
	\dd {\Phi}({\cal N})= \tilde{\mu} \, \dd {\cal N} + \tilde{D} \ \dd W({\cal N}) \,,
\end{align}
where the drift $\tilde{\mu}$ and diffusion $\tilde{D}$ are nearly constant. 
Integrating  the above equation from ${\Phi}_*(k)$ to ${\Phi}_{\rm end}$ 
we obtain 
\begin{equation}
	{\Phi}_{\rm end}({\cal N})-{\Phi}_*= \tilde{\mu} \, {\cal N}+\tilde{D} \int_{0}^{\cal N} \dd W({\cal N}') \, .
\end{equation}
Using the stochastic property of the  Brownian motion such as  the integral \eqref{Ave_StoIntegrals_1}, one obtains
\begin{equation}
	\left\langle {\cal N}^2  \right\rangle= \Big(\dfrac{\tilde{D}}{\tilde{\mu}}\Big)^2 \,.
\end{equation}
Then the curvature power spectrum \eqref{Power2} is the square of the diffusion over drift
\begin{align}
	\label{Power3}
	{\cal P}_\zeta = \Big(\dfrac{\tilde{D}}{\tilde{\mu}}\Big)^2 \,.
\end{align}
In App.~\ref{PowerPDF}, we re-derive the above relation using the PDF method.

Now  considering the Langevin equation \eqref{philong3}, the power spectrum of the curvature perturbation from Eq. (\ref{Power3}) is  obtained to be 
\begin{align}
\label{power1}
{\cal P}_{\zeta}(\varepsilon,\xi) = \dfrac{H^2}{8\pi^2\Mp^2 \epsilon_{_V}}\Big[1+ \dfrac{H^2}{72\sqrt{2}\pi^3\xi^{3/2}} \big(\dfrac{\alpha}{f}\big)^2 \ e^{2\pi \xi} \ \sinh(2\pi \xi) \ \varepsilon^4 \ {\cal G}^2(\varepsilon,\xi)\Big]\,,
\end{align}
where ${\cal G}(\varepsilon,\xi)$ is given by \eqref{g} and is plotted in Fig.~\ref{fig:g}.   The first term above is the contribution from the vacuum fluctuations while the second term is the contribution from the gauge field perturbations. 

In the weak backreaction regime $\xi < \xi_{\rm c}$, with $\xi_{\rm c}$ to be read from Fig.~\ref{fig:ep-xi}, one can eliminate the parameter $\alpha/f$ from the COBE normalization \eqref{calP0}, obtaining
\begin{align}
	\dfrac{\alpha}{f} H_{\rm _{CMB}} \simeq 4\pi \xi_{_{\rm CMB}} \sqrt{{\cal P}_\zeta^{(0)}} \, .
\end{align}
Using this expression, the curvature perturbation power spectrum 
Eq. (\ref{power1}) for $\xi >1$ is written as   
\begin{align}
\label{power11}
{\cal P}_{\zeta}(\xi) \simeq {\cal P}_{\zeta}^{(0)}\Big(1+\dfrac{\sqrt{2\xi}}{18\pi}\varepsilon^4 {\cal P}_{\zeta}^{(0)} \ e^{4\pi \xi} \ {\cal G}^2(\varepsilon,\xi)\Big) \,.
\end{align}
Correspondingly, the fractional correction in power spectrum induced from the gauge field perturbations in the stochastic formalism  is given by
\begin{align}
\label{Deltapower}
\dfrac{\Delta {\cal P}^{\rm Sto}_{\zeta}}{{\cal P}_{\zeta}^{(0)}} \simeq \varepsilon^4 {\cal P}_{\zeta}^{(0)} \ e^{4\pi \xi} \ {\cal G}^2(\varepsilon,\xi) \,.
\end{align}

\begin{figure}[t]
	\centering
	\includegraphics[width=0.8\linewidth]{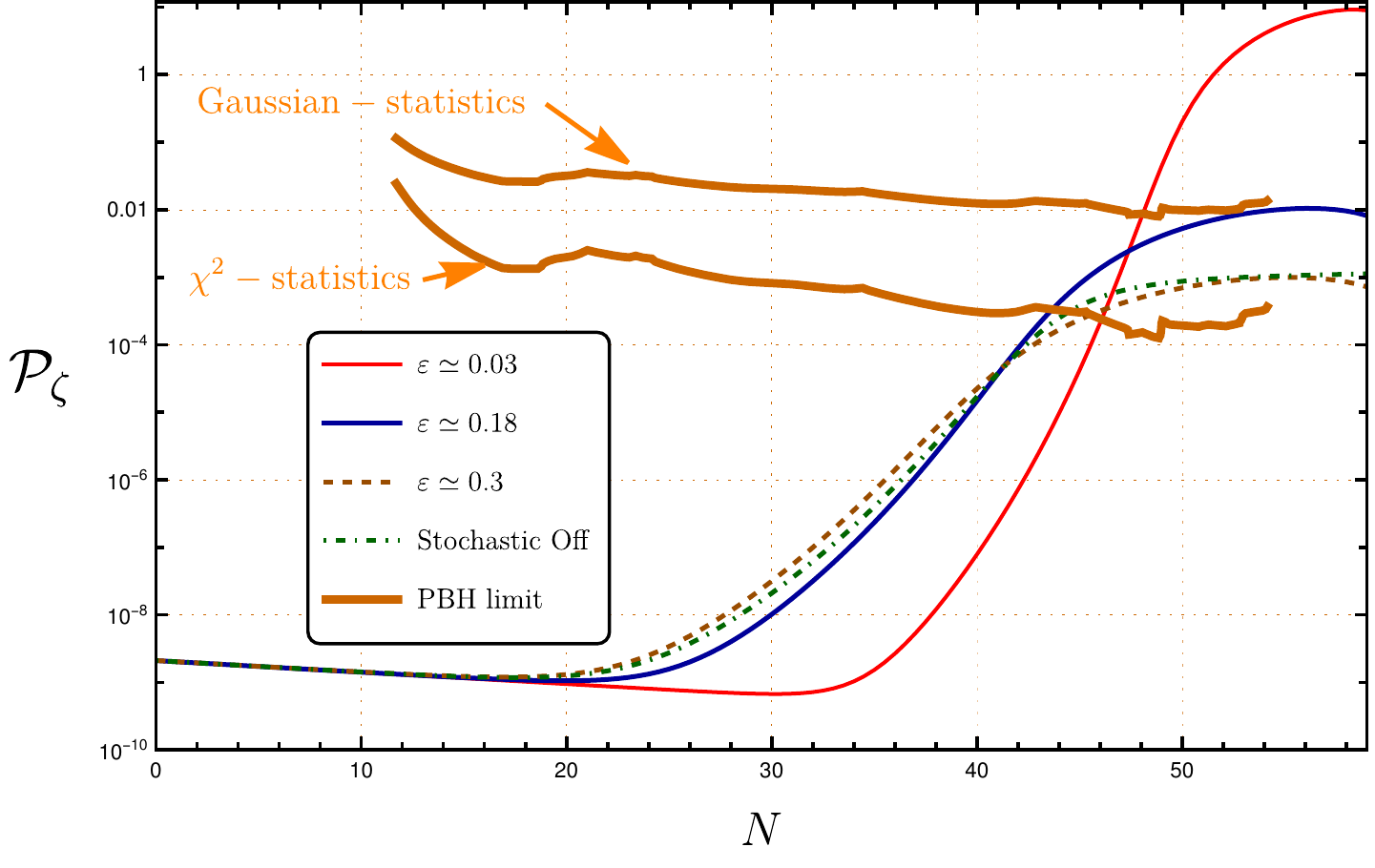}
	\caption{\footnotesize The curvature perturbation power spectrum in axion inflation with the chaotic potential $V(\phi) \propto \phi^2$. The instability parameter at CMB scale, $\xi_{_{\rm CMB}} \simeq 2.2$, is chosen such that the results here can be compared with those of \cite{Linde:2012bt}. The upper thick solid orange lines represent  the bound on the primordial density perturbations to prevent the  overproduction of PBHs for two different statistics: Gaussian (top) and $\chi^2$ (bottom). The result for the case $\varepsilon=0.3$ nearly coincides with the result  in conventional approach with no stochastic noises. The case $\varepsilon=0.03$, which crosses the PBH limit, corresponds to the lower bound of \eqref{epsilon_lower}. Starting with $\xi_{_{\rm CMB}}=2.2$ with $\varepsilon \simeq \{0.03, 0.18, 0.3\}$, the final value of the instability parameter is found to be $\xi(t_e) = \{5.3, 4.2, 3.9\}$ respectively. This shows that the system does not experience any significant backreaction from the gauge field. But, for the case $\varepsilon\simeq 0.03$, the system becomes  non-perturbative with   ${\cal P}_\zeta \gtrsim {\cal O}(1)$ before the end of inflation as seen from the red solid line.}
	\label{fig:Power}
\end{figure}

The above expression should be compared with the corresponding result given in 
Eq. (\ref{DeltapowerBarnaby}) obtained in the conventional approach based on mean-field approximation of tachyonic modes.  
We see that the overall exponential growth $e^{4\pi \xi}$ is the same in both formula which is the hallmark of the curvature perturbations induced from the tachyonic gauge fields perturbations. However, our formula has the stochastic factor $\varepsilon^4$ while  Eq. (\ref{DeltapowerBarnaby}) contains  a numerical suppression $10^{-5}$ which emerged upon approximations employed in obtaining  Eq. (\ref{DeltapowerBarnaby}). This suggests that with $\varepsilon \sim 10^{-1} $ the two methods yield to qualitatively similar results for the induced power spectrum. This is also consistent with the bound obtained in 
Eq. (\ref{bound-epsilon}). Indeed, as discussed in Appendix \ref{EM-noises}, we can take $\varepsilon \sim  e^{-\pi \xi/2} $. To fix the numerical value of $\varepsilon$ one can consider the largest value of  $\xi_{_{\mathrm{CMB}}}$ which the backreactions are under control throughout inflation and then set $\varepsilon \sim  e^{-\frac{\pi}{2} \xi_{_\mathrm{CMB} }}$.

Note that as $\xi$ grows during inflation, one can end up with a situation such that  ${\cal P}_\zeta \gtrsim {\cal O}(1)$ at the end of inflation so the perturbative approximation is violated. In order for the perturbative treatment to be valid during entire period of inflation, one should start with small enough initial value $\xi_{_\mathrm{CMB} }$. 
For example, starting with  $\xi_{_\mathrm{CMB}}=2.2$  and $\varepsilon =0.03$
leads to $\xi{(t_e)} \simeq 5.6$ at the time of end of inflation. This is below $  \xi_{\rm c} \simeq 5.6$ which is obtained from the background constraint but leads to ${\cal P}_\zeta \gtrsim {\cal O}(1)$, invalidating the perturbative treatment. Physically, this originates from the fact 
that the tachyonic growth of the gauge field quanta backreacts on the inflaton field itself by inverse decay: $\delta A + \delta A \rightarrow \delta\varphi$ which causes the enhancement of the scalar power spectra. Note that this is totally different from the backreaction effects arising from the gauge field on the background evolution. These effects can be seen in Fig.~\ref{fig:Power} where the total power spectrum has been plotted vs. the number of e-folds $N$.


\section{PBH Limits on Power Spectrum}
\label{sec:PBH}

In this section, we study the PBHs formation in this setup
 to put  constraints on the model parameters.

A PBH may form in the early universe if there is an enhancement in ${\cal P}_\zeta(k)$ generated during inflation \cite{ivanov1994inflation,Garcia-Bellido:1996mdl,Ivanov:1997ia, Sasaki:2018dmp} on small (sub-CMB) scales.  The small-scale perturbations re-enter the cosmological horizon during radiation era. If these perturbations are large enough in amplitude, they can collapse and form a PBH of mass similar to the horizon mass (see, for example, Refs. \cite{Germani:2018jgr,Musco:2018rwt} for more details on the criterion for formation). We have obtained an enhancement in 
${\cal P}_\zeta$  given in Eq. \eqref{power11} due to tachyonic growth of the gauge field perturbations towards the end of inflation.  Therefore, the probability of PBH formation in this scenario is not negligible  \cite{Lin:2012gs, Linde:2012bt}. 
Correspondingly, the enhanced power may  lead to the overproduction of PBH which can overclose the universe. This  can also be used to put limits on the model parameters.

A PBH will form if at horizon re-entry ($k = a H$) the amplitude of the smoothed density contrast $\delta$ is large enough. The classical PBH formation criterion in the radiation-dominated epoch is given by \cite{carr1974black},
\begin{align}
	\label{delta-c}
	\delta > \delta_{\rm c} 
	\,,
\end{align}
where $\delta$ is the smoothed density contrast at horizon crossing, $k = a H$. The probability of having $\delta > \delta_{\rm c}$ corresponds to the
fraction of space $\beta$ that can collapse to form horizon-sized
black holes. The parameter $\beta$ represents the mass fraction (the energy density fraction) of PBHs at the time of formation,
\begin{align}
	\label{beta-rho}
	\beta \equiv \dfrac{\rho_{\rm PBH}}{\rho_{\rm tot}} \bigg\vert_{t_{\rm f}} 
	= \Omega_{\rm PBH}
	\ \left(
	\dfrac{H_0}{H_{\rm f}}
	\right)^2 \left(
	\dfrac{a_{\rm f}}{a_0}
	\right)^{-3} \,,
\end{align}
where $\Omega_{\rm PBH}$ is the density parameter of  PBHs at present. The subscribes ``$0$'' and ``${\rm f}$'' denote the values evaluated at the present and at the time of formation $t_{\rm f}$, respectively. 

The PBH mass can be approximated by the horizon mass, $M_H \equiv(4\pi/3)\rho_{\rm f} H_{\rm f}^{-3}$, with $\rho_{\rm f}$ being the total energy density of the Universe at $t_{\rm f}$. We then find
\begin{align}
	M_{_{\rm PBH}} = \gamma \ M_{H} 
	= \dfrac{4\pi \gamma \Mp^2}{H_{\rm f}} \,,
\end{align}
in which  $\gamma$ is a correction factor evaluated as $\gamma= 3^{-3/2}\simeq0.2$ by a simple analytic
calculation for the collapse in the radiation dominated era \cite{carr1975primordial}. 

On the other hand, having the PDF of $\delta$, denoted by $f_\delta(x)$, the mass fraction is given by 
\begin{align}
	\label{beta_PDF}
	\beta = \int_{\delta_{\rm c}}^{\delta_{\rm max}} {\rm d}x \ f_\delta(x) \,,
\end{align}
where $\delta_{\rm max}$ is the maximum value of the density perturbation at the horizon crossing for PBH formation. The PDF of $\delta$ is simply related to the PDF of primordial curvature perturbations as follows. The comoving density perturbation $\delta$ is related to the Bardeen potential $\Psi$ in Fourier space through the relation
\begin{align}
	\label{delta-Bardeen}
	\delta_k = -\dfrac{2}{3}\left(
	\dfrac{k}{aH}
	\right)^2 \Psi_k \,.
\end{align}
For the super-horizon modes, $\Psi_k \simeq -\frac{2}{3}\zeta_k$, the criterion \eqref{delta-c} can be translated to a lower bound for the curvature perturbation, which is
\begin{align}
	\zeta > \zeta_{\rm c} \simeq \dfrac{9}{4}\delta_{\rm c} \,.
\end{align}
If we assume the PBH formation threshold $\delta_{\rm c} \approx 0.45$ \cite{Musco:2008hv} ($\delta_{\rm c} \approx 1/3$\cite{carr1974black}), then $\zeta_{\rm c} \sim 1$ ($\zeta_{\rm c} \sim 0.75$). Therefore, a PBH is formed when a curvature mode re-enters the horizon during radiation era when its amplitude is above $\zeta_{\rm c}$~. The probability of this event can be read from Eq.~\eqref{beta_PDF} when it is written in the following form
\begin{align}
	\label{beta_PDF_zeta}
	\beta = \int_{\zeta_{\rm c}}^{\zeta_{\rm max}} {\rm d}x \ f_\zeta(x) \,,
\end{align}
where $f_\zeta$ is the PDF of primordial curvature perturbations. 

Conventionally, it has been assumed that $f_\zeta$ obeys a $\chi^2$-statistics \cite{Linde:2012bt,Bugaev:2013vba,Bugaev:2012ai}. This assumption is based on the fact that the curvature perturbation is the sum of a vacuum modes plus a part sourced by the gauge modes. Since the vacuum term is always negligibly small for PBH formation, one only needs to consider the formation due to the source term which originates from the convolution of two Gaussian modes. The non-Gaussianity of fluctuations $\delta \phi$, described by Eq. \eqref{Eom_Linde} in conventional approach,  arises just from the particular solution \eqref{phi-J} which is bilinear in the gauge field. Therefore in this context the PDF $f_\zeta$ follows a $\chi^2$-statistics.

Contrary to the above view, taking into account the stochastic noises, the scalar fluctuations $\phi_{\boldsymbol k}$ is now  described by Eq. \eqref{KG_phik} which is linear in quantum gauge field perturbations. Specifically, in the right hand side of Eq. \eqref{KG_phik} as the source term, we have the product of a classical long mode and a quantum short mode so the corresponding statistics is expected to be Gaussian due to the quantum short mode fluctuations. To support this conclusion,  in App.~\ref{appendix:PDF} we have calculated the PDF of curvature perturbation using the Langevin equation \eqref{philong3}. We have shown that, to a good accuracy, $f_\zeta$ follows a Gaussian distribution, 
\begin{align}
	\label{f_zeta}
	f^{\rm G}_\zeta(x;\sigma_{\zeta}) &=  \dfrac{e^{-\frac{x^2}{2\sigma_\zeta^2}}}{\sqrt{2\pi \sigma_\zeta^2}} \, ,
\end{align}
in which $\sigma_\zeta$ is the variance of the curvature fluctuations.

A PBH forms  when   $\zeta(k_{\mathsf N}) \gtrsim \zeta_{\rm c}$, where we recall
that $k_{\mathsf N}$ indicates the wavenumber corresponding to the mode that has left the horizon ${\mathsf N}$ $e$-folds before the end of inflation, ${\mathsf N}=\abs{N_{\rm tot}-N}$. Very naively, the variance is considered as $\sigma_\zeta^2 \equiv \langle \zeta(k_{\mathsf N})^2 \rangle = {\cal P}_\zeta (k_{\mathsf N})$. To derive the relation between the number of $e$-folds $\mathsf N$ and the PBH mass $M_{_{\rm PBH}}$ that can be formed from this mode, we assume the universe is radiation dominated right after the end of inflation (i.e. assuming an instant reheating). In this case, the black hole mass can be estimated as \cite{Garcia-Bellido:1996mdl}
\begin{align}
	M_{_{\rm PBH}} \simeq 10 \, {\rm gr} \, \big(\dfrac{\gamma}{0.2}\big) \Big(\dfrac{10^{-6} M_{\rm Pl} \, H_{\rm end}}{{H_{\mathsf N}^2}}\Big) \, 
	e^{2{\mathsf N}} \, ,
\end{align}
where $H_{\rm end}$ and $H_{\mathsf N}$ are the Hubble rates at the time of end
of inflation and when
the mode $k_{\mathsf N}$ exits the horizon during inflation, respectively. Above, we have normalized the scale factor at the end of inflation to unity and used $k_{\mathsf N}=e^{-\mathsf N} H_{\mathsf N}$~. We then obtain $\sigma_\zeta^2 \simeq  {\cal P}_\zeta (M_{_{\rm PBH}})$  for the variance of PDF in terms of PBH mass. 

To be more precise, in order to calculate the probability of PBH formation 
we need the PDF of the smoothed $\zeta$-field, $\zeta_R$, where $R$ is the smoothing radius. The smoothing effects come only through the variance $\sigma_\zeta$, while the shape of the PDF is the same as in Eq. \eqref{f_zeta}. Let us  introduced the smoothed variance $\sigma_{_{\rm PBH}}$ as
\begin{align}
	\sigma_{_{\rm PBH}}^2 \equiv \langle \zeta_R^2 \rangle = \dfrac{16}{81} \int_{0}^{\infty} {\rm d}\ln k \ (kR)^4 \ {\cal P}_\zeta(k) \ \tilde{W}^2(kR) \, ,
\end{align}
where $R^{-1}=a_{\rm f} H_{\rm f}$ is the comoving scale at $t_{\rm f}$ and $\tilde{W}(kR)$ is a Fourier transform of the Gaussian window function, $\tilde{W}(kR)=e^{-k^2 R^2}$. Putting all  together, one finds that the fraction $\beta$ of the Universe which goes into PBH of mass scale $M_{_{\rm PBH}}$ at the formation epoch is given by
\begin{align}
	\label{beta:Gaussian}
	\beta(M_{_{\rm PBH}}) &= 2\gamma \int_{\zeta_{\rm c}}^{\zeta_{\rm max}} \ f^{\rm G}_\zeta(x;\sigma_{_{\rm PBH}}) \ \dd x \, ,
	\nonumber
	\\
	&\simeq \gamma \ {\rm Erfc}\Big( \dfrac{\zeta_{\rm c}}{\sqrt{2}\sigma_{_{\rm PBH}}} \Big) \, ,
	\nonumber
	\\
	&\simeq \gamma \ \sqrt{\dfrac{2}{\pi}} \dfrac{\sigma_{_{\rm PBH}}}{\zeta_{\rm c}} \ \exp(-\frac{\zeta_{\rm c}^2}{2\sigma_{_{\rm PBH}}^2}) \,;
	\hspace{1cm}
	\sigma_{_{\rm PBH}}^2 = \sigma_{_{\rm PBH}}^2(M_{_{\rm PBH}}) \,,
\end{align}
where the factor 2 comes from the Press-Schechter theory and ${\rm Erfc}(x)$ is the complementary error function. We have assumed $\zeta_{\rm max} \gg \zeta_{\rm c} \gg \sigma_\zeta$ in the second and last equations. The last expression is a consequence of the asymptotic expansion of ${\rm Erfc}(x)$ for $x \gg 1$. 

Using the above relation, one can compute the fraction of PBHs against the total DM density at the present given by \cite{Sasaki:2018dmp}
\begin{align}
	f(M_{_{\rm PBH}}) \equiv \dfrac{\Omega_{\rm PBH}}{\Omega_{\rm DM}} \simeq 1.52 \times 10^{8}  \ \beta(M_{_{\rm PBH}})  \ \left(
	\dfrac{\gamma}{0.2}
	\right)^{1/2} \left(
	\dfrac{g^*_{\rm f}}{106.75}
	\right)^{-1/4} \left(
	\dfrac{M_{_{\rm PBH}}}{M_\odot}
	\right)^{-1/2} \,,
\end{align}
in which $g^*_{\rm f}$ is the number of relativistic degrees of freedom when PBHs form and $M_\odot \simeq 2 \times 10^{33}~{\rm gr}$ is the solar mass. For the power spectrum given by \eqref{power11} where the peak of the power spectrum
is located in the  last ten $e$-folds, the fraction $f$ is significant only for PBHs with the mass less than $ \lesssim 10^{11} {\rm gr}$. Since the PBHs with mass $ \lesssim 10^{15} {\rm gr}$ have evaporated by the present epoch via  Hawking radiation, PBH of cosmological interests can not be generated in the simplest model of  axion inflation.  In Sec.~\ref{sec:Discussion} we discuss a variant of axion inflation  which can generate more massive PBHs to be relevant for cosmological 
purposes such as  for dark matter or in GWs studies.   

\begin{figure}[t!]
	\centering
	\includegraphics[width=0.8\linewidth]{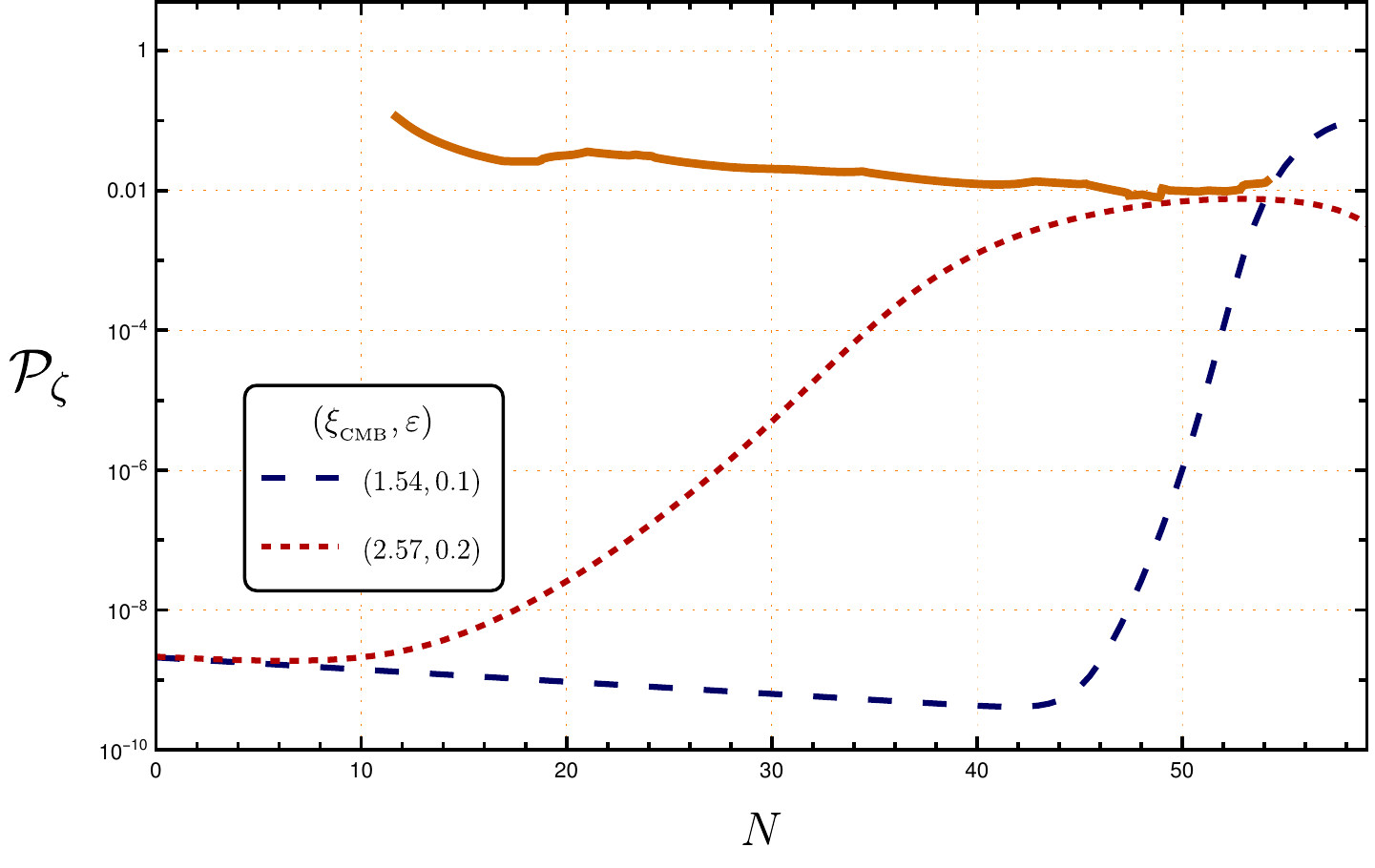}
	\caption{\footnotesize The enhancement of curvature power spectrum for quadratic  potential. The upper thick solid orange line represents the PBHs bound on the primordial density perturbations for the  Gaussian scalar perturbations with 
$\zeta_{\rm c} = 1$. {The parameters have been chosen  such that the PBHs bound is not violated. For this potential, the blue dashed curve corresponds to $\xi_{_{\rm CMB}} \simeq 1.54$ with the cutoff parameter $\varepsilon \simeq 0.1$ coming from the lower bound in Eq. \eqref{epsilon_lower}. In this case, the final value for instability parameter at the end of inflation  is $4.64$ which is below the  backreaction bound $\xi_{\rm c}=4.97$~. The red dotted curve corresponds to $\varepsilon \simeq 0.2$ and $\xi_{_{\rm CMB}} \simeq 2.57$ with the final value 
$\xi(t_e) \simeq 4.14$ which is  smaller than $\xi_{\rm c}=4.53$~. As seen from both cases, the perturbative scheme is valid in which $\calP_\zeta < 1$}. }
	\label{fig:Power16100}
\end{figure}

Here, therefore,  we only deal with the PBH constraints on the model parameters. Specifically, the PBH bounds can be translated to the bounds on $\xi_{_{\rm CMB}}$ and $\alpha/f$ \cite{Linde:2012bt,Bugaev:2013fya}. There are observational constraints on $f$ coming from the non-detection of PBHs. We have used these constraints from Ref. \cite{Garcia-Bellido:2016dkw} and references therein. Thus, for a given mass of PBH, the constraints on $f$ can be interpreted as the  constraints on $\beta$ \cite{Garcia-Bellido:2016dkw}. The limits on $\beta$ can also be  translated into the upper bounds on the primordial scalar density perturbations as a function of  $\mathsf N$. In Fig.~\ref{fig:Power}, we have presented the upper bound by assuming a constant Hubble rate $H_{\mathsf N} \simeq H_{\rm end} =10^{-6} M_{\rm Pl}$ during inflation for two different statistics of the induced primordial scalar perturbations.

If the induced scalar modes obey a $\chi^2$-statistics, as in conventional treatment studied in previous literature, 
then the mass fraction can be estimated as
\begin{align}
	\label{beta: chi^2}
	\beta_{\chi^2} \simeq \gamma \, {\rm Erfc}\Big[ \big( \dfrac{1}{2}+\dfrac{\zeta_{\rm c}}{\sqrt{2}\sigma_{_{\rm PBH}}} \big)^{\frac{1}{2}}  \, \, \Big] \,,
\end{align}
which significantly tightens the limit on the scalar power with respect to Gaussian statistics \eqref{beta:Gaussian}. Therefore, we conclude that  taking into account the stochastic noises along with the Gaussian distribution of primordial perturbations relax the constraints from  the overproduction of PBHs on 
the model parameters. However,  the significant enhancement of power spectrum at the end of inflation due to stochastic noises must be considered especially when one chooses a  very small values of $\varepsilon$.  

Here, we present the PBHs constraints for two particular potentials, $V(\phi) \propto \phi^p$ for $p=1$ and $p=2$.  Compared to the previous works \cite{Bugaev:2013fya,Linde:2012bt}, we can translate the constraints on the overproduction of PBHs  to  constraints on $\xi_{_{\rm CMB}}$ at CMB scales. 
The authors of \cite{Bugaev:2013fya} obtained the constraint on the value of instability parameter $\xi_{_{\rm CMB}}$ for the linear (quadratic) potential to be $\xi_{_{\rm CMB}} \lesssim 1.65$~($\xi_{_{\rm CMB}} \lesssim 1.75$). To obtain these results, they assumed that the $\zeta$-field has a $\chi^2-$distribution\footnote{These values are relaxed  by about three percents if we consider $\chi^2_2-$distribution~\cite{Bugaev:2013fya}.}. The constraints obtained in \cite{Bugaev:2013fya,Linde:2012bt} are based on the fact that there are no black hole bounds for the last six $e$-folds of inflation,  ${\mathsf N} \lesssim 6$. Turning on the stochastic noises, however, the distribution of $\zeta$-field is Gaussian. This relaxes  the upper bounds on $\xi_{_{\rm CMB}}$ compared to the 
 $\chi^2$-distribution. In Fig.~\ref{fig:Power16100}, we have  presented the results for the quadratic potential.  We have obtained the constraints $\xi_{_{\rm CMB}} \lesssim 2.57$ for the quadratic potential. For a linear potential one obtains $\xi_{_{\rm CMB}} \lesssim 2.50$. The results show that the stochastic noises shift the previous bounds on the instability parameter towards  larger values by about fifty percents. The main reason for this difference is that we have a  Gaussian distribution of curvature perturbation in the stochastic formalism. 
 In addition, we see that the enhancement in power spectrum towards the last six e-folds of inflation  is stronger  in the  stochastic approach.

\begin{figure}[t!]
	\centering
	\includegraphics[width=0.7\linewidth]{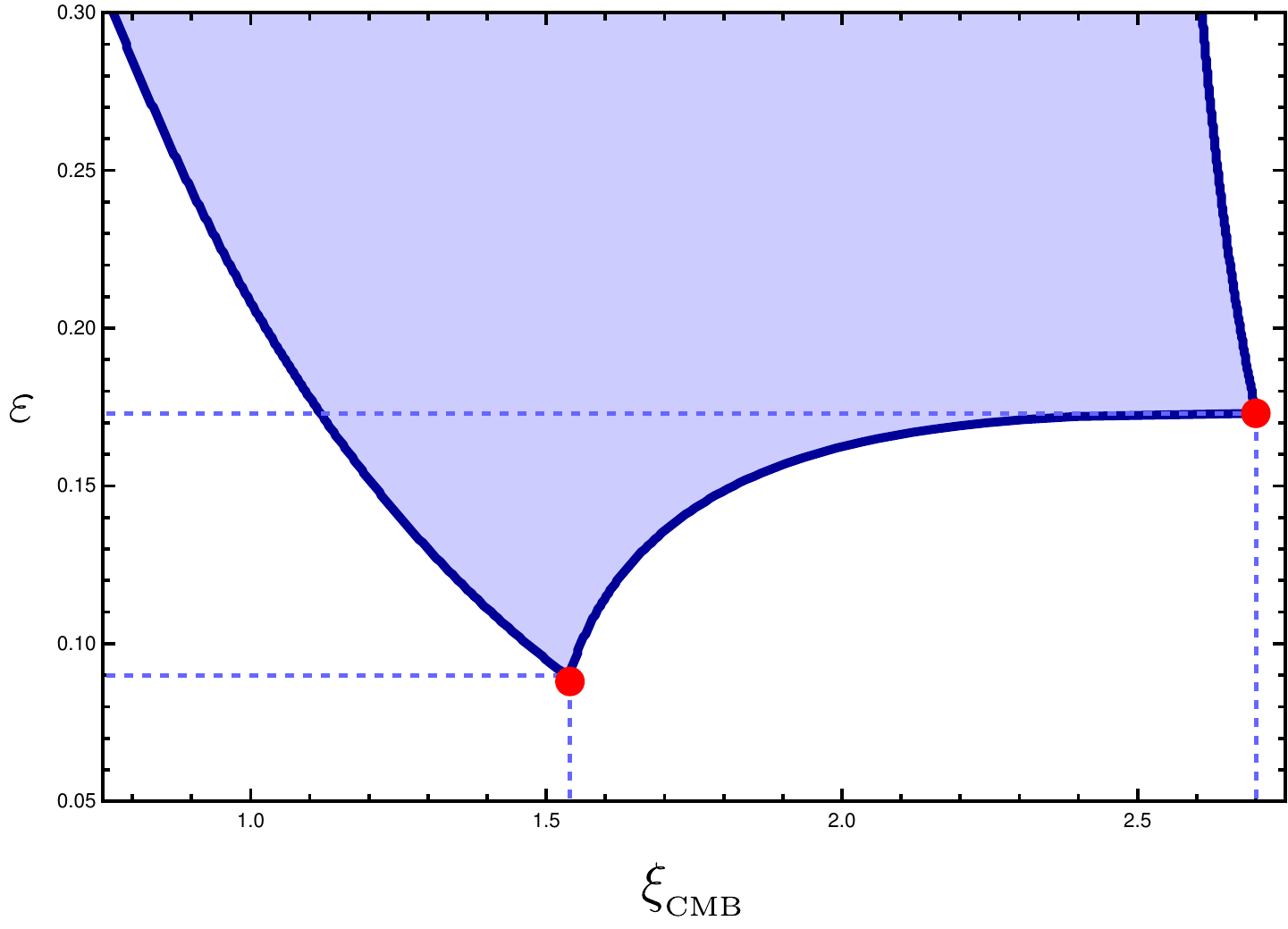}
	\caption{\footnotesize {The allowed parameter space for the cutoff parameter $\varepsilon$ vs. the instability parameter $\xi_{_\mathrm{CMB} }$. The shaded 
blue area satisfy all four requirements indicated at the end of this subsection, i.e.
the PBHs bound, no strong backreactions on background dynamics, the power 
spectrum to be perturbative and the COBE normalizations on the CMB scales. 
The two red dots have the coordinates $(1.54, 0.09)$ and $(2.7, 0.173)$. The instability parameter for the former point, $\xi_{_{\rm CMB}} \simeq 1.5$, is near the result  obtained in the absence of stochastic noises as studied in Refs.\cite{Bugaev:2013fya,Linde:2012bt}. However, the cutoff parameter $\varepsilon \simeq 0.17$ for the latter point is around the intersection point (red dot) in Fig.~\ref{fig:ep-xi} where the backreaction effects at the background level 
in stochastic formalism coincides with those of conventional method. Imposing the 
additional condition $ \varepsilon \lesssim m/H$ from the scalar noise  \eqref{epsilon_lower}, we conclude that  the value $\varepsilon \simeq {\cal O}(0.1)$ is the typical acceptable value of the cutoff parameter.}}
	\label{fig:ep_xi_CMB}
\end{figure}

The comparison between the stochastic approach and the conventional method becomes more transparent if we investigate the parameter space of $(\xi_{_\mathrm{CMB} },\varepsilon)$ in the presence of perturbations. 
In Fig.~\ref{fig:ep_xi_CMB}, we have presented the allowed parameter space of $(\xi_{_\mathrm{CMB} },\varepsilon)$ for the quadratic potential while the following  conditions are satisfied: 
\begin{enumerate}
	\item The PBH bounds on the power spectrum, arising from the Gaussian distribution of curvature perturbation, are satisfied.
	\item The tachyonic growth of the gauge fields does not induce strong backreactions on the background  inflaton dynamics. In other words, the allowed  parameter space (blue area) shown in Fig.~\ref{fig:ep-xi} is chosen.
	\item The induced curvature perturbations are perturbatively under control, 
	i.e. ${\cal P}_\zeta < {\cal O}(1)$ throughout inflation.
	\item The COBE normalization for the power spectrum of curvature perturbation \eqref{calP0} has been imposed on CMB scales. 
\end{enumerate}
After imposing the above constraints, we see from  Fig.~\ref{fig:ep_xi_CMB} 
that the cutoff parameter  $\varepsilon$
typically is at the order  $\varepsilon \sim {\cal O}(0.1)$
as we mentioned before. Also note that large value of $\varepsilon$, while 
acceptable in Fig.~\ref{fig:ep_xi_CMB}, are not allowed  as it will be in conflict with  the upper bound of  \eqref{epsilon_lower}, $ \varepsilon \lesssim m/H$, coming from the scalar noise.

\section{Summary and Discussions}
\label{sec:Discussion}

In this paper, we have revisited the model of axion inflation by taking into account the stochastic effects of electromagnetic noises.  Because of the parity violating interaction, one of the polarization of the gauge field perturbations become tachyonic inducing large curvature perturbations. The amplitude of the induced power spectrum is controlled by the  instability parameter $\xi$ which evolves adiabatically during inflation.  

We have  derived the associated Langevin equations for the electric and magnetic fields, given respectively by  Eqs. \eqref{Elong2} and \eqref{Blong2}. The latter has the form of an OU process while the former is in the form of Vasicek SDE. The 
main feature of these two kinds of SDE is that they describe  a \textit{mean-reverting} process during which the fields settles into their equilibrium states. This  property prevents them from decaying and also from experiencing a very large tachyonic instability.  In addition, the local equilibrium of electromagnetic fields 
protects the inflaton field from the tachyonic growth of the gauge fields towards the end of inflation.  The stochastic noises relax the bounds on the instability parameter $\xi$ before the system enters the strong backreaction regime. 
In the conventional approach studied in previous literature  one usually estimates a background value for the electromagnetic fields by calculating the cumulative effects of tachyonic modes. However,  in the stochastic approach, we study the evolution of coarse grained electromagnetic fields taking into account the stochastic noise arising from the UV modes. The strength of backreactions in two approaches are qualitatively  the same for the cutoff parameter 
of $\varepsilon \sim {\cal O}(0.1)$.  This value of $\varepsilon$ is supported from 
various constraints imposed both at the background and perturbation levels.  However, having a Gaussian distribution for $\zeta$ in stochastic formalism can distinguish these two approaches from each other. 

We have studied  the Langevin equation of the inflaton field and calculated the curvature perturbation power spectrum  induced by the  gauge field perturbations. We have shown that  the distribution of gauge field 
curvature perturbation follows a  Gaussian statistics and have studied  the PBH formation  in the presence of the stochastic noises. As $\xi$ evolves adiabatically and  the curvature perturbations is amplified only towards the last 5-10 e-folds of inflation, the produced PBH are  light and are evaporated  via Hawking radiation. Imposing the PBH constraints  we have found the upper bounds $\xi_{_{\rm CMB}} \lesssim 2.50$ and $\xi_{_{\rm CMB}} \lesssim 2.57$ for the linear and quadratic potentials respectively. Consequently, the bounds on $\xi$ are shifted by  more than fifty percents towards larger values.  The main reason for this difference is that we have a  Gaussian distribution for  the induced curvature perturbations.

Motivated from the above results, specially on the roles of the stochastic effects in  estimating the backreactions, it would be interesting to explore the stochastic approach in other scenarios such as in models where the axion field is not the inflaton. Also one can look at tensor perturbations in this setup in the presence of  stochastic noises, since the tachyonic gauge fields affect not only the scalar perturbations but also the gravitational waves. Another good question is how stochastic noises affect the non-Gaussianity of primordial curvature perturbation. In addition, having the solution \eqref{Beq} for the stationary state of the magnetic field, one can look for the amplitude of  the primordial magnetic fields on large scales generated in this setup. We leave these issues to future works. Moreover, in a work in progress, we would like to  investigate a model in which axion field experiences a period of ultra-slow-roll (USR) phase during inflation. In that setup, the instability parameter 
$\xi$ falls off rapidly  during USR regime while there exists an enhancement in power spectrum as in conventional USR phase. PBHs formation in this USR axion setup shows a few interesting features.

\vspace{1cm}
{\bf Acknowledgments:} We would like to thank Sina Hooshangi for useful comments in numerical calculations. H. F. and A. T. are partially supported by the ``Saramadan" federation of Iran. 
\vspace{0.5cm}

\appendix
\section{Scalar Mode Function}
\label{scalar-mode}

The evolution of scalar mode function is given by
\begin{eqnarray}
\label{KG_k}
\ddot{\varphi}_{\boldsymbol k} + 3H\dot{\varphi}_{\boldsymbol k}+\left(\dfrac{k^2}{a^2}+m^2
\right)\varphi_{\boldsymbol k} = J_{\boldsymbol k} \,,
\end{eqnarray}
in which $m^2=V_{,_{\phi\phi}}$ determines the mass of the scalar field and $J_{\boldsymbol{k}}$ is a source term, given by
\begin{align}
J_{\boldsymbol k} =  \dfrac{\alpha}{f}\big(
{{\boldsymbol E}}_l
.{{\boldsymbol B}}_{\boldsymbol k}+
{{\boldsymbol B}}_l
.{{\boldsymbol E}}_{\boldsymbol k} \big) \,.
\end{align}
There are two differences between \eqref{KG_k} and the Fourier transform of \eqref{Eom_Linde} relating to the friction and the source terms. The friction term in the conventional approach has an additional contribution from gauge quanta while in stochastic approach the friction is controlled by the usual $3 H$ factor. 
Moreover, in stochastic approach we have a linear term for the quantum 
mode (i.e. only one ${{\boldsymbol E}}_{\boldsymbol k}$ or ${{\boldsymbol B}}_{\boldsymbol k}$ in the source accompanied by the classical terms ${{\boldsymbol E}}_l$  and ${{\boldsymbol B}}_l$)  but the corresponding  source term in  
Eq. \eqref{Eom_Linde} appears as a non-linear convolution in Fourier transform of \eqref{Eom_Linde} which made the calculations more difficult.

The solution of \eqref{KG_k} consists of two parts: the first part is the homogeneous solution and the second part is the particular solution which is due to the source. Schematically, we denote these two contributions  as
\begin{align}
\label{decom_Phi}
\phi_{\boldsymbol k} = \underbrace{\varphi_{\boldsymbol k}^{\rm vac}}_{\rm homogeneous} + \underbrace{\varphi_{\boldsymbol k}^{ J}}_{\rm particular} \,. 
\end{align}
Physically, the homogeneous solution corresponds to the vacuum fluctuations while the particular solution arises due to inverse decay processes $ A_\lambda + A_\lambda \rightarrow \varphi_{\boldsymbol k}^J$. Remember that the homogeneous solutions of \eqref{KG_k} are given by the well-known result 
\begin{equation}
\label{vac_mode}
\varphi_k^{\rm vac}(\eta) = -i \frac{H\sqrt{-\pi \eta^3}}{2}  \ H^{(1)}_{\nu}(-k\eta), \hspace{5mm}\nu \cong \frac{3}{2} + \mathcal{O}(\epsilon_H,\eta_H) \, ,
\end{equation}
where we have assumed the mass of the scalar field is very small compared to the Hubble scale, $V_{,\phi \phi} \ll H$. 

To obtain the particular solution, it is more convenient to rewrite Eq. \eqref{KG_k} in terms of conformal time for the new variable $v_k(\eta) \equiv a(\eta) \ \varphi_k(\eta)$,
\begin{eqnarray}
&&  \left[\partial_\eta^2 + k^2 + a^2 m^2 - \frac{a''}{a} \right] v_k(\tau) = a^3(\eta) \ J_{\bf k}(\eta) \,.
\end{eqnarray}
The homogeneous solution of the above equation is then given by $v_k^{\rm vac}(\eta)=a(\eta) \ \varphi_k^{\rm vac}(\eta)$ when \eqref{vac_mode} is used. Using the Green  function, satisfying
\begin{align}
\left[\partial_\eta^2 + k^2 + m^2 a^2  - \frac{a''}{a} \right]   G_k(\eta,\eta') = \delta \left( \eta - \eta' \right) \,,
\end{align}
the particular solution $v_k^J(\eta)$ can be obtained. We employ the vacuum modes $v_k^{\rm vac}(\eta)$ in the retarded Green function,
\begin{equation}
\label{green}
G_k(\tau,\tau') = i \ \Theta(\eta-\eta') \  \bigg[ v_k^{\rm vac}(\eta) \ v_k^{{\rm vac}\, \star}(\eta') -  v_k^{{\rm vac}\, \star}(\eta) \ v_k^{\rm vac}(\eta') \bigg] \,,
\end{equation}
to obtain $v_k^{J}(\eta)=a(\eta) \ \varphi_k^{J}(\eta)$. Putting things together, one  obtains
\begin{equation}
\label{J_mode}
\varphi_k^{J}(\eta) =\dfrac{1}{a(\eta)} \int_{-\infty}^{0} \ \dd \eta'  \ G_k(\eta,\eta') \ a^3(\eta') \  J_{\bf k}(\eta') \,.
\end{equation}

It is worth mentioning that the homogeneous and the particular solutions are statistically independent of each other. In fact, the homogeneous solution $\varphi_k^{\rm vac}$ can be expanded in terms of the creation $\hat{b}^\dagger_{\boldsymbol{k}}$ and annihilation $\hat{b}_{\boldsymbol{k}}$ operators associated with the inflaton vacuum fluctuations, while the particular solution $v_k^J$ can be expanded in terms of the ladder operators $\hat{a}^{\lambda}_{\boldsymbol{k}}, \hat{a}^{\lambda' \dagger}_{\boldsymbol{k}}$ associated with the gauge fields. As seen from Eq. \eqref{commut_ab},  these two sets of operators commute with one another.

\section{Noise Correlations}
\label{noise}

In this appendix, we  derive the explicit forms of the quantum noises Eqs. \eqref{noisephi}-\eqref{noiseB}. For noise correlations of the scalar fields 
we have \cite{Grain:2017dqa}
\begin{align}
\label{phi_correlation}
\left \langle
\sigma_{_\phi}(t_1,\bfx) \ \sigma_{_\phi}(t_2,\bfx)
\right \rangle
&= \dfrac{1}{6\pi^2} \dfrac{{\rm d}k_{_c}^3}{{\rm d}t}  \ \big|\varphi_{_{k_c}}(t_1)\big|^2 \  \delta(t_1-t_2)\,,
\\
\label{tau_correlation}
\left \langle
\tau_{_\phi}(t_1,\bfx) \ \tau_{_\phi}(t_2,\bfx)
\right \rangle
&= \dfrac{1}{6\pi^2} \dfrac{{\rm d}k_{_c}^3}{{\rm d}t} \ \big|\dot{\varphi}_{_{k_c}}(t_1)\big|^2 \  \delta(t_1-t_2)\,.
\end{align}
While for the helical electromagnetic fields, $X=E, B$, we obtain\cite{Talebian:2021dfq}
\begin{align}
\label{X_sigma}
\left \langle
\sigma^{_X}_i(t_1,\bfx) \ \sigma^{_X}_j(t_2,\bfx)
\right \rangle
&= \dfrac{1}{18\pi^2} \dfrac{{\rm d}k_{_c}^3}{{\rm d}t} \ \sum_{\lambda}^{}\big|X_\lambda(t_1,k_c)\big|^2 \ \delta_{ij} \ \delta(t_1-t_2) 
\\
\label{X_tau}
\left \langle
\tau^{_X}_i(t_1,\bfx) \ \tau^{_X}_j(t_2,\bfx)
\right \rangle
&= \dfrac{1}{18\pi^2} \dfrac{{\rm d}k_c^3}{{\rm d}t} \ \sum_{\lambda}^{}\big|\dot{X}_\lambda(t_1,k_c)\big|^2 \delta_{ij} \ \delta(t_1-t_2)
\end{align}
For the non-helical electromagnetic fields, the above relations are consistent with the results of \cite{Talebian:2019opf,Talebian:2020drj}. In what follows, we 
calculate simple relations for the above quantum noises and show that they could be expressed via the classical white noises.

\subsection{Electromagnetic Noises}
\label{EM-noises}

For the electromagnetic fields, we use the definitions \eqref{E-B-Def} for the gauge field mode function $A_\lambda$ \eqref{Alambda} to obtain mode functions $E_\lambda$ and $B_\lambda$. After expanding  $E_\lambda$ and $B_\lambda$ around $k_{_c}=\varepsilon a H$ where $\varepsilon \rightarrow 0$ and choosing the leading term, we find
\begin{align}
\label{BE_lambda}
B_\lambda(k_c) = i \dfrac{H^2 \ e^{\frac{\pi \xi \lambda}{2}}}{\sqrt{2} \ \xi \ k_c^{3/2} \ \Gamma(-i\xi\lambda)}\varepsilon^2 \,,
\hspace{1cm}
E_\lambda(k_c) = -i \dfrac{\sqrt{2}H^2 \ e^{\frac{\pi \xi \lambda}{2}}}{k_c^{3/2} \ \Gamma(-i\xi\lambda)}\varepsilon^2 \ln\varepsilon \,.
\end{align}
Applying the above expressions into Eqs. \eqref{X_sigma} and \eqref{X_tau}, we obtain \footnote{$|\Gamma(i\xi \lambda)|^2= \dfrac{\pi}{\xi \lambda \sinh(\pi\xi\lambda)}$}
\begin{align}
\label{Cor1}
\left \langle
\sigma^{_E}_i(N_1) \ \sigma^{_E}_j(N_2)
\right \rangle
&=%
\dfrac{H^6}{3\pi^3}\xi \sinh(2\pi\xi) \
\varepsilon^4 (\ln\varepsilon)^2 \ \delta_{ij} \ \delta(N_1-N_2) \,,
\\
\label{Cor2}
\left \langle
\sigma^{_B}_i(N_1) \ \sigma^{_B}_j(N_2)
\right \rangle
&=%
\dfrac{H^6}{12\pi^3} \dfrac{\sinh(2\pi\xi)}{\xi} \
\varepsilon^4 \ \delta_{ij} \ \delta(N_1-N_2) \,,
\end{align}
and
\begin{align}
\label{Cor3}
	\left \langle
	\tau^{_E}_i(N_1) \ \tau^{_E}_j(N_2)
	\right \rangle
	&=%
	\dfrac{4H^8}{3\pi^3}\xi \sinh(2\pi\xi) \
	\varepsilon^4 (\ln\varepsilon)^2 \ \delta_{ij} \ \delta(N_1-N_2) \,,
	\\
	\label{Cor4}
	\left \langle
	\tau^{_B}_i(N_1) \ \tau^{_B}_j(N_2)
	\right \rangle
	&=%
	\dfrac{3H^8}{\pi^3} \xi \ \sinh(2\pi\xi) \
	\varepsilon^6 (\ln\varepsilon)^2 \ \delta_{ij} \ \delta(N_1-N_2) \,.
\end{align}
Hereafter we have used number of $e$-fold, ${\rm d}N \equiv H{\rm d}t$ as the clock. 

The quantum noises $\sigma$ and $\tau$ become classical when commute each other, $[ \sigma^{_X} , \tau^{_X} ] = 0$. For both the electric and magnetic fields we find 
\begin{align}
\label{Cor5}
	\left \langle
	[\sigma^{_X}_i(N_1) , \tau^{_X}_j(N_2)]
	\right \rangle
	&=%
	\dfrac{i \ H^7}{3\pi^2} \ \varepsilon^5 \ \delta_{ij} \ \delta(N_1-N_2) \,.
\end{align}
Therefore, by considering $\varepsilon^5 \rightarrow 0$, the quantum noises $\sigma$ and $\tau$ become classical noises. On the other hand, we can not take 
$\varepsilon$ arbitrarily close to zero as then the amplitude of the electric and magnetic noises in Eqs. (\ref{Cor1}) and (\ref{Cor2}) go to zero. As in the case of scalar field (see next subsection), we demand that $\varepsilon \rightarrow 0$ in  such a way that $\sinh(2\pi\xi) \varepsilon^4 \sim 1$. This in turn fixes the scales of 
$\varepsilon$ to be $\varepsilon \propto  e^{-\pi \xi/2}$. For $\xi \sim 2$ we typically have  $\epsilon \lesssim 10^{-1}$.

We define a three dimensional (3D) Wiener process ${\bf W}$ associated with a 3D normalized white noises $\boldsymbol{\Xi}$ via
\begin{align}
\label{Wiener}
\dd {\bf W}(N) \, \equiv \, {\bf \Xi}(N) \, \dd N \,,
\end{align}
where
\begin{align}
\langle \boldsymbol{\Xi}(N) \rangle = 0 \,,
\hspace{3cm}
\langle \Xi_i(N_1) \, \Xi_j(N_2) \rangle = \delta_{ij} \,\, \delta(N_1-N_2)\,.
\end{align}
Now, one can express the electric and magnetic noises in terms of the normalized white noise as
\begin{align}
\label{sigmaE}
\boldsymbol{\sigma}_{_X}(N) \equiv \Mp H^2 \, D_{_X} \,\, \boldsymbol{\Xi}(N) 
\end{align}
where
\begin{align}
\label{DX}
D_{_X} &= \dfrac{H}{2\Mp} \, \sqrt{\dfrac{\sinh(2\pi\xi)}{3\pi^3\xi}} \, \varepsilon^2 \times
\begin{cases}
2 \xi \abs{\ln\varepsilon}  \,, \qquad & X = E
\\
\\
1 \,,\qquad & X = B
\end{cases}
\,.
\end{align}

One can search for the  direction dependency of the electric and the magnetic noises. Assume the wave number $\hat{\boldsymbol k}$ and the polarization vectors $\boldsymbol{e}_\lambda(\hat{\boldsymbol{k}})$ are given by
\begin{align}
\label{k}
\hat{\boldsymbol k} &=
\left(
\sin\theta \cos\phi ,\ \sin\theta \sin\phi ,\ \cos\theta
\right) \,,
\\
\label{e}
\boldsymbol{e}_\lambda(\hat{\boldsymbol{k}}) &= \dfrac{1}{\sqrt{2}}
\left(
\cos\theta \cos\phi-i\lambda \sin\phi,\ \cos\theta \sin\phi +i\lambda \cos\phi,\ -\sin\theta
\right) \,.
\end{align}
We can calculate the Cartesian components of the electric and magnetic noises by considering $\bfx=r \ \hat{k}$. It is easy to show that
\begin{align}
\sigmaE_i (t,\bfx) &= \hat{i} \ . \ \bdsigmaE(t,\bfx) \propto \int_{\phi=0}^{2\pi}{\rm d}\phi \ \sum_\lambda \hat{i} \ . \  e_i^{\lambda}(\hat{\boldsymbol{k}}) \ \left[ E_\lambda(t,k) \,\hat{a}^\lambda_{\boldsymbol{k}} +  E_{\lambda}^{*}(t,k)\, \hat{a}^{\lambda \dagger}_{-\boldsymbol{k}} \right] = 0
\\
\sigmaE_j (t,\bfx) &= \hat{j} \ . \ \bdsigmaE(t,\bfx) \propto \int_{\phi=0}^{2\pi}{\rm d}\phi \ \sum_\lambda \hat{j} \ . \  e_i^{\lambda}(\hat{\boldsymbol{k}}) \ \left[ E_\lambda(t,k) \,\hat{a}^\lambda_{\boldsymbol{k}} +  E_{\lambda}^{*}(t,k)\, \hat{a}^{\lambda \dagger}_{-\boldsymbol{k}} \right] = 0
\\
\sigmaE_k (t,\bfx) &= \hat{k} \ . \ \bdsigmaE(t,\bfx) \ne 0
\end{align}
These calculations show that the electric and magnetic noises have aligned along the $\bfx$ direction, 
\begin{align}
\label{alignX}
\bdsigmaB(t,\bfx) \propto \bfx \,,
\hspace{3cm}
\bdsigmaE(t,\bfx) \propto \bfx \,.
\end{align}
Here we use the fact that for any $\lambda$-dependent function $g_\lambda$, one has 
\begin{align}
\sum_{\lambda = \pm}  \ g_\lambda \ e_i^{\lambda}(\hat{\boldsymbol{k}})~e_j^{\lambda *}(\hat{\boldsymbol{k}}) &=\dfrac{1}{2}\sum_{\lambda = \pm}  \ g_\lambda\left(
\delta_{ij}-\hat{k}_i \hat{k}_j + i\lambda \ f_{ij}(\theta,\phi) \, ,
\right)
\end{align}
where $f_{ij}$ is an anti-symmetric function, $f_{ij}=-f_{ji}$, and is given by
\begin{align}
f_{21}(\theta)=\cos\theta \,,
\hspace{8mm}
f_{32}(\theta,\phi)=\sin\theta \cos\phi\,,
\hspace{8mm}
f_{13}(\theta,\phi)=\sin\theta \sin\phi \,.
\end{align}

\subsection{Scalar Noises}
\label{scalar-noises}

To compute the correlation function of the stochastic noises of the scalar field we consider the decomposition of \eqref{decom_Phi} with the homogeneous and particular solutions \eqref{vac_mode} and \eqref{J_mode} respectively. Since the homogeneous solution $\varphi_k^{\rm vac}$ is expanded in terms of $\hat{b}^\dagger_{\boldsymbol{k}}$ and $\hat{b}_{\boldsymbol{k}}$ 
which are independent of the operators $\hat{a}^{\lambda}_{\boldsymbol{k}}$ and $\hat{a}^{\lambda' \dagger}_{\boldsymbol{k}}$, according to \eqref{commut_ab}, it is more convenient to split the quantum noise of the scalar field into two parts:
\begin{align}
\label{noise_split}
\left \langle
\sigma_{_\phi}(t_1,\bfx) \ \sigma_{_\phi}(t_2,\bfx)
\right \rangle
&\equiv
\left \langle
\sigma^{\rm vac}(t_1,\bfx) \ \sigma^{\rm vac}(t_2,\bfx)
\right \rangle
+
\left \langle
\sigma^J(t_1,\bfx) \ \sigma^J(t_2,\bfx)
\right \rangle \,,
\\
\left \langle
\tau_{_\phi}(t_1,\bfx) \ \tau_{_\phi}(t_2,\bfx)
\right \rangle
&\equiv
\left \langle
\tau^{\rm vac}(t_1,\bfx) \ \tau^{\rm vac}(t_2,\bfx)
\right \rangle
+
\left \langle
\tau^J(t_1,\bfx) \ \tau^J(t_2,\bfx)
\right \rangle \,.
\end{align}

In general, the quantum noises $(\sigma , \tau)$ are not independent of the mass, $m^2 \equiv V_{,\phi \phi}$. For a light scalar field, i.e. $0< m^2 \ll H^2$,  with the mode functions \eqref{vac_mode},  it is well-known that\cite{Starobinsky:1986fx,Sasaki:1987gy, Nambu:1988je, Nakao:1988yi}
\begin{align}
\label{cor1}
\left \langle
\sigma^{\rm vac}(N_1,\bfx) \ \sigma^{\rm vac}(N_2,\bfx)
\right \rangle
&= \varepsilon^{2m^2/3H^2} \dfrac{H^4}{4\pi^2} \ \delta(N_1-N_2) \,,
\\
\label{cor2}
\left \langle
\tau^{\rm vac}(N_1,\bfx) \ \tau^{\rm vac}(N_2,\bfx)
\right \rangle
&= \varepsilon^{2m^2/3H^2}\bigg(\dfrac{m^2}{3H^2}+\varepsilon^2\bigg)^2\dfrac{H^6}{4\pi^2} \ \delta(N_1-N_2) \,,
\\
\label{Anticor1}
\left \langle
\{ 
\sigma^{\rm vac}(N_1,\bfx) , \tau^{\rm vac}(N_2,\bfx) 
\}
\right \rangle
&= -\varepsilon^{2m^2/3H^2}\bigg(\dfrac{m^2}{3H^2}+\varepsilon^2\bigg)\dfrac{H^5}{4\pi^2} \ \delta(N_1-N_2) \,,
\\
\label{cor3}
\left \langle
[\sigma^{\rm vac}(N_1,\bfx) , \tau^{\rm vac}(N_2,\bfx) ]
\right \rangle
&= i \dfrac{H^5}{4\pi^2} \ \varepsilon^3 \ \delta(N_1-N_2) \,,
\end{align}
where $[ .. \,, .. ]$ and $\{ .. \,, .. \}$ denote the commutator and anti-commutator operators. 

From the above equations we find that the quantum nature of $(\sigma^{\rm vac},\tau^{rm vac})$ becomes negligible  if  $\exp(-3H^2/m^2) \ll \varepsilon^2 \ll 1$. In particular, if $m^2=0$, the momentum noise $\tau^{\rm vac}$ can be neglected by choosing $\varepsilon$ sufficiently small. While for $m^2 \neq 0$, the amplitudes of the noises become independent of $\varepsilon$ for $\varepsilon \ll m^2/3H^2$. Therefore, it is safe to choose $\varepsilon$ \cite{Sasaki:1987gy, Nambu:1988je, Nakao:1988yi}
 from the range $\exp(-3H^2/2 m^2) < \varepsilon < m/H$. 

Equations \eqref{cor1}-\eqref{cor3} show that we have the relation
\begin{align}
	\label{tau-sigma}
	\tau^{\rm vac} \simeq -\dfrac{m^2}{3H} \,  \sigma^{\rm vac} \, .
\end{align}
Actually, the above relation can be directly derived from the definitions \eqref{noisephi} and \eqref{noisepi}.

For the source part, we have to compute the correlation of $\varphi_k^J$ and therefore the correlation of $J_{\boldsymbol k}$. It is a straightforward calculation to show that the momentum noise $\tau^J$, arising from the particular solution \eqref{J_mode}, has the same relation \eqref{tau-sigma} with $\sigma^J$ so
we do not consider it anymore.

\begin{figure}[t!]
	\centering
	\includegraphics[width=0.75\linewidth]{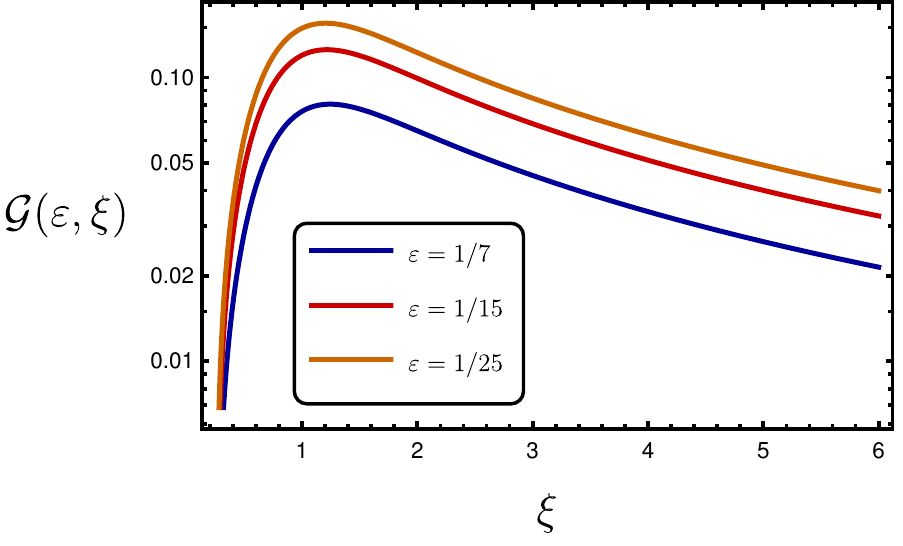}
	\caption{\footnotesize The function ${\cal G}(\varepsilon,\xi)$ defined in \eqref{g} and appearing in the power spectrum of scalar field and the amplitude of the correlation of stochastic noises.}
	\label{fig:g}
\end{figure}

Using Eqs. \eqref{k}, \eqref{e} and the tachyonic mode function of the gauge field \eqref{amplified_A}, we obtain
\begin{align}
&\langle J_{{\boldsymbol k}_1}(\eta_1) \ J_{{\boldsymbol k}_2}(\eta_2) \rangle
= (2\pi)^3 \
\left(\dfrac{\alpha}{f}\right)^2 \ \sin^2\theta  \
 \delta^3\left({\boldsymbol k}_1  - {\boldsymbol k}_2 \right)
\nonumber
\\
&\times \bigg[ {\cal E}_{\rm eq} B_+(\eta_1)B_+(\eta_2)
+
{\cal E}_{\rm eq}{\cal B}_{\rm eq} \bigg(E_+(\eta_1)B_+(\eta_2)+B_+(\eta_1)E_+(\eta_2)\bigg)
+
{\cal B}_{\rm eq} E_+(\eta_1)E_+(\eta_2) \bigg] \, .
\end{align}
To proceed with the computation, we define
\begin{align}
{\cal F}_n(\xi) \equiv \int_{1/8\xi}^{2\xi} \dd x \ \left(\dfrac{\sin x}{x}-\cos x\right) x^{-n/4} \ e^{-2\sqrt{2\xi x}} \,.
\end{align}
One obtains
\begin{align}
\left \langle
\sigma^J(t_1,\bfx) \ \sigma^J(t_2,\bfx)
\right \rangle = \dfrac{H^5\sqrt{2}}{576\pi^5 \xi^{3/2}} \left(\dfrac{\alpha}{f}\right)^2 \varepsilon^4 \ e^{2\pi \xi} \ \sinh(2\pi \xi) \ {\cal G}^2(\varepsilon,\xi) 
\, \delta(t_1-t_2)  \, ,
\end{align}
in which 
\begin{align}
\label{g}
{\cal G}(\varepsilon,\xi) &\equiv  2\xi \ \abs{\ln\varepsilon} \ {\cal F}_3(\xi) + \dfrac{1}{4} {\cal F}_7(\xi)-\sqrt{2\xi}{\cal F}_5(\xi) \,.
\end{align}
This function is plotted in Fig.~\ref{fig:g} for $0.25<\xi<10$ and three different values of $\varepsilon$.

Taking all these together, finally we obtain
\begin{align}
\left \langle
\sigma_{_\phi}(N_1) \ \sigma_{_\phi}(N_2)
\right \rangle
=
\left(\dfrac{H^2}{2\pi}\right)^2
\bigg[
1+\varepsilon^4 \ \dfrac{H^2\sqrt{2}}{12^2\pi^3\xi^{3/2}} \left(\dfrac{\alpha}{f}\right)^2 \ e^{2\pi \xi} \ \sinh(2\pi \xi) \ {\cal G}^2(\varepsilon,\xi)
\bigg] \delta(N_1-N_2) \, .
\end{align}
The first term above is the contribution of the vacuum scalar modes while the second term represents the contributions from the  gauge fields perturbations  through the inverse decay process:  ${{\boldsymbol E}}_l
.{{\boldsymbol B}}_{\boldsymbol k}+
{{\boldsymbol B}}_l
.{{\boldsymbol E}}_{\boldsymbol k} \rightarrow \phi_{\boldsymbol k}$.

The same as for the electromagnetic case, we introduce a Wiener process $W$ associated with a normalized white noises $\Xi$ via
\begin{align}
\label{Wiener-scalar}
\dd W(N) \, \equiv \, \Xi(N) \, \dd N \,,
\end{align}
\begin{align}
\langle \Xi(N) \rangle = 0 \,,
\hspace{3cm}
\langle \Xi(N_1) \, \Xi(N_2) \rangle =  \delta(N_1-N_2)\, .
\end{align}
Now, one can rewrite the scalar noises in terms of the normalized white noise as
\begin{align}
\label{sigmaPhi}
\sigma_{_\phi}(N) \equiv H \, D_{_\phi} \,\, \Xi(N) \,,
\end{align}
where
\begin{align}
\label{DPhi}
D_{_\phi} &=
\dfrac{H}{2\pi}
\bigg[
1+\varepsilon^4 \ \dfrac{H^2\sqrt{2}}{12^2\pi^3\xi^{3/2}} \left(\dfrac{\alpha}{f}\right)^2 \ e^{2\pi \xi} \ \sinh(2\pi \xi) \ {\cal G}^2(\varepsilon,\xi)
\bigg]^{1/2}
\,.
\end{align}
We use the above relation for the amplitude of the scalar noise in the main draft.

\section{Power Spectrum from PDF}
\label{PowerPDF}
In this appendix we justify the relation \eqref{Power3} used for the power spectrum by means of probability distribution function. To this end we start by the following  Langevin equation 
\begin{equation}\label{langevin}
	\phi({\cal N})-\phi_0= \mu \, {\cal N} + D \,  W({\cal N})\,,
\end{equation}
where $\mu$ and $D$ are the constant drift and diffusion coefficients respectively.
We study the first boundary crossing as studied in \cite{Firouzjahi:2018vet} and show that if one of the barriers is far enough from the other one, then the power spectrum is simply given as
\begin{equation}
	\mathcal{P}_\zeta=\frac{d\left<\delta\mathcal{N}^2\right>}{d\left<\mathcal{N}\right>}=\frac{D^2}{\mu^2} \, .
\end{equation}

Now suppose that we have two barriers $\phi_\pm$ with the initial condition set at $\phi_0$ with $p_+$ ($p_-$) the conditional probability of hitting $\phi_+$ ($\phi_-$)
before $\phi_-$ ($\phi_+$). Assuming  $|\phi_+|<<|\phi_-|$ then one expects that the probability $p_+$($p_-$) that the field hits $\phi_+$($\phi_-$) earlier than $\phi_-$($\phi_+$) is equal to 1(0).  If one can set $p_-\phi_-=0$, which should be justified,  then using the  fact that $\big<W(\mathcal{N})^2)\big>=\big<\mathcal{N}\big>$\cite{Firouzjahi:2018vet} we have
\begin{equation}
	\left<\phi-\phi_0\right>=p_+(\phi_+-\phi_0)+p_-(\phi_--\phi_0)=\phi_+-\phi_0=\mu\left<\mathcal{N}\right> \, .
\end{equation}
And then one obtains
\begin{equation}\label{averge}
	\left<\mathcal{N}\right>=\frac{\phi_+-\phi_0}{\mu}
\end{equation}
Moreover, we have
\begin{equation}\label{phi2}
	\left<(\phi-\phi_0)^2\right>=(\phi_+-\phi_0)^2=\mu^2\left<\mathcal{N}^2\right>+2\mu D\left<W(\mathcal{N})\mathcal{N}\right>+D^2\left<W(\mathcal{N})^2\right> \, .
\end{equation}
Now one can write 
\begin{equation}
	\begin{split}
		\left<W(\mathcal{N})\mathcal{N}\right>&=\Big<W(\mathcal{N})(\frac{\phi-\phi_0-D W(\mathcal{N})}{\mu})\Big>=\\&p_+\frac{\phi_+-\phi_0}{\mu}\left<W(\mathcal{N})|\phi=\phi_+\right>+p_-\frac{\phi_--\phi_0}{\mu}\left<W(\mathcal{N})|\phi=\phi_-\right>-\frac{D}{\mu}\left<\mathcal{N}\right>
	\end{split}
\end{equation}
As $p_-\phi_-=0$ then one can set the second term in the above equation equal to zero and simply set $\left<W(\mathcal{N})|\phi=\phi_+\right>=\left<W(\mathcal{N})\right>=0$. Then the second term in \eqref{phi2} is zero and we have
\begin{equation}\label{n2}
	\left<(\phi-\phi_0)^2\right>=(\phi_+-\phi_0)^2=\mu^2\left<\mathcal{N}^2\right>-D^2\left<\mathcal{N}\right> \, ,
\end{equation}

Then one can easily read $\left<\mathcal{N}^2\right>$ as 
\begin{equation}
	\left<\mathcal{N}^2\right>=\frac{(\phi_+-\phi_0)^2+D^2\, \frac{(\phi_+-\phi_0)}{\mu}}{\mu^2}
\end{equation}

Using Eqs. \eqref{n2} and \eqref{averge}
one gets
\begin{equation}\label{power23}
	\mathcal{P}_\zeta=\frac{d\left<\delta\mathcal{N}^2\right>}{d\phi_0}\frac{d\phi_0}{d\left<\mathcal{N}\right>}=\frac{D^2}{\mu^2} \, .
\end{equation}
Now we should justify our use of $p_-\phi_-=0$. To show that this holds for our Langevin equation \eqref{langevin} we use the probability distribution function of $\phi$ with two barriers $\phi_\pm$ and show that the behaviour of moments $\mathcal{N}$ is the same as what we obtained with stochastic calculus. It can be shown that the conditional probability distribution functions corresponding to  
Eq. \eqref{langevin} is as follows \cite{karatzas2014brownian}
\begin{align}
	\begin{split}
		f_+(N) &= \frac{1}{\sqrt{2\pi D^2 N^3}}\sum_{n=-\infty}^{n=\infty}(2n(\phi_+-\phi_-)+(\phi_0-\phi_-))\exp(\frac{\mu(\phi_--\phi_0)}{D^2}-\frac{\mu^2N}{2D^2})
		\\
		&\times\exp(-\frac{(2n(\phi_+-\phi_-)+(\phi_0-\phi_-))^2}{2D^2N})
	\end{split}
	\\
	\begin{split}
		f_-(N) &= \frac{1}{\sqrt{2\pi D^2 N^3}}\sum_{n=-\infty}^{n=\infty}(2n(\phi_+-\phi_-)+(\phi_0-\phi_+))\exp(\frac{\mu(\phi_+-\phi_0)}{D^2}-\frac{\mu^2N}{2D^2})
		\\
		&\times\exp(-\frac{(2n(\phi_+-\phi_-)+(\phi_0-\phi_+))^2}{2D^2N})
	\end{split}
\end{align}
By $f_+$($f_-$)  as the conditional probability one can easily determine the moments of $\big<N_{+}\big>$($\big<N_{-}\big>$)  by the condition that that $\phi_+$($\phi_-$) is hit earlier than $\phi_-$($\phi_+$). Note that $f_\pm$ are not normalized and their integral yield $p_\pm$.
Now one can write the moments of these two conditional distributions as follows:
\begin{equation}
	M_{\pm}(s)=\int_{-\infty}^{\infty}\exp(sN)f_\pm(N) \, ,
\end{equation}
where $s<0$. Note that having the moments at hand one can easily calculate different moments of $\big<\mathcal{N}_\pm\big>$ by taking the derivative of $M_\pm$ with respect to $s$. In other words, we have $\lim_{s\rightarrow0} M_\pm^{n}(s)=p_\pm\big<\mathcal{N}_\pm^n\big>$ .One can then show that 
\begin{equation}
	M_\pm(s)=\frac{\exp({\frac{\phi _\pm \left(\sqrt{\mu^2+2 D^2 s}+\mu\right)-\mu \phi _0}{D^2}}) \left[ \exp({\frac{\phi _0 \sqrt{\mu^2+2 D^2 s}}{D^2}})-\exp({\frac{\left(2 \phi _\mp-\phi _0\right) \sqrt{\mu^2+2 D^2 s}}{D^2}})\right]}{\exp({\frac{2 \phi _\pm \sqrt{\mu^2+2 D^2 s}}{D^2}})-\exp({\frac{2 \phi _\mp \sqrt{\mu^2+2 D^2 s}}{D^2}} )}
\end{equation}
Now one can easily calculate the time average and the squared time average using the following relations:
\begin{align}\label{times}
	\begin{split}
		\left<\mathcal{N}\right> &= \lim_{s\rightarrow0}(M'_-(s)+M'_+(s))\\&=\frac{e^{-\frac{2 \mu \phi _0}{D^2}} \left[\phi _- \left(e^{\frac{2 \mu \left(\phi _-+\phi _+\right)}{D^2}}-e^{\frac{2 \mu \left(\phi _-+\phi _0\right)}{D^2}}\right)-\phi _+ e^{\frac{2 \mu \left(\phi _-+\phi _+\right)}{D^2}}\right]}{\mu \left(e^{\frac{2 \mu \phi _-}{D^2}}-e^{\frac{2 \mu \phi _+}{D^2}}\right)}\\&
		+\frac{\left(\phi _+-\phi _0\right) e^{\frac{2 \mu \left(\phi _++\phi _0\right)}{D^2}}+\phi _0 e^{\frac{2 \mu \left(\phi _-+\phi _0\right)}{D^2})}}{{\mu \left(e^{\frac{2 \mu \phi _-}{D^2}}-e^{\frac{2 \mu \phi _+}{D^2}}\right)}}
	\end{split}
	\\
	\label{times2}
	\left<\mathcal{N}^2\right> &= \lim_{s\rightarrow0}(M''_-(s)+M''_+(s)) \, ,
\end{align}
where we have used the fact that $\big<\mathcal{N}^n\big>=p_+\big<\mathcal{N}_+\big>+p_-\big<\mathcal{N}_-\big>$. We have not represented the explicit form of  $\left<\mathcal{N}^2\right>$ here, as it is complicated. It is interesting to see the  behaviour of $\left<\mathcal{N}\right>$ and $\left<\mathcal{N}^2\right>$ for $\phi_-\rightarrow-\infty$. In this limit one can show that
\begin{equation}
	\lim_{\phi_-\rightarrow-\infty}\left<\mathcal{N}\right>=\frac{\phi_+-\phi_0}{\mu},
\end{equation}
\begin{equation}
	\lim_{\phi_-\rightarrow-\infty}\left<\mathcal{N}^2\right>=\frac{(\phi_+-\phi_0)^2+D^2\frac{(\phi_+-\phi_0)}{\mu}}{\mu^2},
\end{equation}
consistent with what we obtained by the stochastic calculus. 
One can show that in this limit we have
\begin{equation}
	\lim_{\phi_-\rightarrow-\infty} p_-\phi_-=\lim_{\phi_-\rightarrow-\infty}\lim_{s\rightarrow0}\phi_-M_-(s)=0.
\end{equation}
So our primary assumption is justified.

\section{PDF from Langevin Equation}
\label{appendix:PDF}
In this appendix we estimate the PDF of the inflaton in the axion model up to leading and next to leading order. Our method is based on Volterra equation which is discussed in \cite{Pattison:2021oen, buonocore1990two}. As we will see the probability density is approximately Gaussian in the drift dominated regime in the axion model.

To this end we start  by the following Langevin equation:
\begin{equation}\label{axionlang}
\frac{d\phi}{dN}=A(\phi)B(N)+C(\phi)D(N)\xi(N),
\end{equation}
where $A(\phi)$ and $c(\phi)$ are  functions of $\phi$ and $B(N)$ and $D(N)$ are time dependent functions with slow varying derivatives which are at the order of the slow-roll parameters. Note that if we set $B(N)$ and $D(N)$ equal to unity  then we reproduce the ordinary Langevin equation in the slow-roll inflation.

The main idea of Volterra equation approach is to transform the time dependent drift and diffusion of the Langevin equation into the equation of 
a time dependent barrier. So we may look for a function like $z(N)$ which satisfies the following equation:
\begin{equation}\label{brownian}
\frac{dz(N)}{dN}=\xi(N) \,,
\end{equation}
which is pure Brownian motion. Comparing Eq. \eqref{brownian} with Eq.\eqref{axionlang} one finds that
\begin{equation}
z(N)=\int \frac{d\phi}{D(N)C(\phi)}-\int \frac{A(\phi)B(N)}{D(N)C(\phi)}dN \,.
\end{equation}
As a consequence the barriers are now transformed into
\begin{equation}
z_-(N)=\frac{1}{D(N)}\int \frac{d\phi}{C(\phi)}\Big|_{\phi_-}-\int \frac{A(\phi_-)}{C(\phi_-)}\frac{B(N)}{D(N)}dN \,,
\end{equation}
\begin{equation}
z_+(N)=\frac{1}{D(N)}\int \frac{d\phi}{C(\phi)}\Big|_{\phi_+}-\int \frac{A(\phi_+)}{C(\phi_+)}\frac{B(N)}{D(N)}dN \,.
\end{equation}
Note that we are in the case that one of barriers, i.e the initial condition of the field, is playing the role of the reflective barrier and the other one is the absorbing one which is set at the end of inflation. Hence we set $\phi_+=\phi_e$ and $\phi_-=\phi_0$. Now in general one can show that  the PDF of the first time hitting $z_-$ ($z_+$) without before hitting $z_+$ ($z_-$) are given by the following two integral equations\cite{Pattison:2021oen,buonocore1990two}:
\begin{equation}\label{Fminus}
\begin{split}
f_-^{(0)}(N|z_{in},N_{in})=&\Psi_-(N|z_{in},N_{in})-\int^N_{N_{in}}dN'[f_-^{(0)}(N'|z_{in},N_{in})\Psi_-(N|z_-(N'),N')\\&+f_+^{(0)}(N'|z_{in},N_{in})\Psi_-(N|z_+(N'),N')] \,,
\end{split}
\end{equation}
and 
\begin{equation}\label{Fplus}
\begin{split}
f_+^{(0)}(N|z_{in},N_{in})=&-\Psi_+(N|z_{in},N_{in})-\int^N_{N_{in}}dN'[f_-^{(0)}(N'|z_{in},N_{in})\Psi_+(N|z_-(N'),N')\\&+f_+^{(0)}(N'|z_{in},N_{in})\Psi_+(N|z_+(N'),N')],
\end{split}
\end{equation}
where $N_{in}$ and $z_{in}$ denote the initial time and initial value of $z$ respectively. Moreover,  $\Psi_\pm(N|z_{in},N_{in})$ is defined as follows
\begin{equation}
\Psi_\pm(N|z_{in},N_{in})=(z'_\pm(N)-\frac{z_\pm(N)-z_{in}}{N-N_{in}})f(z_\pm,N|z_{in},N_{in}),
\end{equation}
with
\begin{equation}
F(z,N|z_{in},N_{in})=\frac{e^{-\frac{(z-z_{in})^2}{2(N-N_{in})}}}{\sqrt{2\pi(N-N_{in})}} .
\end{equation}
Note that the superscript $(0)$ means that the field has never  bounced the reflective barrier. In general the time distribution function is given as
\begin{equation}
f_{-}(N|z_{in},N_{in})=\sum_{n=0}^\infty  f_{-}^{(n)}(N|z_{in},N_{in}).
\end{equation}
However as the other terms are highly suppressed in the drift dominated regime we will take the first term as the approximate solution. 

Now we  determine $f_-(z,N|z_{in},N_{in})$ up to next to leading order. As we are in the drift dominated regime then one expects that at leading order the PDF behaves as a Dirac delta function:
\begin{equation}
f_-^{LO}(N)=\delta(N-N_{cl}(z_{in},N_{in})),
\end{equation}
where $N_{cl}$ is the classical number of e-folds 
while $f_+^{LO}(N)=0$. 

Substituting these two PDFs into \eqref{Fplus} and \eqref{Fminus} one obtains
\begin{equation}
\begin{split}
f_-^{NLO,(0)}(N|N_{in},z_{in})&=\Big(z'_-(N)-\frac{z_-(N)-z_{in}}{N-N_{in}}\Big)\frac{e^{-\frac{(z-z_{in})^2}{2(N-N_{in})}}}{\sqrt{2\pi(N-N_{in})}}-\\&\Big(z'_-(N)-\frac{z_-(N)-z_{in}}{N-N_{cl}}\Big)\frac{e^{-\frac{(z-z_{in})^2}{2(N-N_{cl})}}}{\sqrt{2\pi(N-N_{cl})}}\theta(N-N_{cl}) \,.
\end{split}
\end{equation}
By expanding the above expression around $N_{cl}$ one gets
\begin{equation}
f_-^{NLO,(0)}(N|N_{in},z_{in})\simeq  \frac{-3 z_{in}e^{-9 z_{in}^2\frac{(N-N_{cl})^2}{2(N_{cl}-N_{in})}}}{\sqrt{2\pi(N_{cl}-N_{in})}} \,,
\end{equation}
which is Gaussian as promised. 

\small
\bibliography{references}
\bibliographystyle{JHEPNoTitle} 

\end{document}